\documentclass[12pt,a4paper]{article}
\usepackage[T1]{fontenc}
\usepackage[utf8]{inputenc}
\usepackage{authblk}
\usepackage{bbding} 
\usepackage{ulem}
\usepackage{soul}

\usepackage{amsmath}
\usepackage{graphicx,float,color}
\usepackage{hyperref}
\hypersetup{colorlinks = true, linkcolor = blue, urlcolor  = blue, citecolor = blue, anchorcolor = blue}
\usepackage{multirow}
\usepackage[numbers,sort&compress]{natbib}
\usepackage{braket}
\usepackage{amsfonts}
\usepackage{amssymb} 
\usepackage{epic}
\usepackage{eepic}
\usepackage{epsfig}
\usepackage{latexsym}
\usepackage{color}
\usepackage{url}  
\usepackage{caption}
\usepackage[caption=false]{subfig}
\usepackage{float}
\usepackage{bm}
\usepackage{geometry}
\usepackage{slashed}
\usepackage[detect-all]{siunitx}

\def\Mp{m_{\mathrm{Pl}}}
\def\lp{\ell_{\mathrm{Pl}}}

\def\Mmin{M_{\mathrm{min}}}
\def\hMmin{\hat{M}_{\mathrm{min}}}

\usepackage{lipsum}
\makeatletter
\def\ps@pprintTitle{%
 \let\@oddhead\@empty
 \let\@evenhead\@empty
 \def\@oddfoot{}%
 \let\@evenfoot\@oddfoot}
\makeatother
\usepackage{etoolbox}
\begin{document}
\title{\textbf{On thermodynamics of compact objects
 }}    
\author[a,b]{Ufuk Aydemir\thanks{uaydemir@metu.edu.tr} }
\author[b]{Jing Ren\thanks{renjing@ihep.ac.cn}}
%\author[2]{Author D\thanks{D.D@university.edu}}
\affil[a]{\normalsize Department of Physics, Middle East Technical University, Ankara 06800, T\"urkiye}
\affil[b]{\normalsize Institute of High Energy Physics, Chinese Academy of Sciences, Beijing 100049, China}
\maketitle

%\date{\today}
\vspace{-0.8cm}
\begin{abstract}
With the recent progress in observations of astrophysical black holes, it has become more important to understand in detail the physics of strongly gravitating horizonless objects. If the objects identified in the observations are indeed horizonless and ultracompact, high curvature effects may become important, and their explorations may be intimately related to new physics beyond General Relativity (GR). In this paper, we revisit the concept of statistical thermodynamics in curved spacetime, focusing on self-gravitating compact systems without event horizons. In the literature, gravitational field equations are in general assumed \textit{a priori} in the thermodynamic treatment, which may lead to difficulties  
for theories of modified gravity, given the more complicated structure of field equations.
Here, we consider
%on the curved spacetime effects 
thermodynamic behavior of the matter source, instead of the physical mass, hence avoiding the explicit input of field equations in the derivation of thermodynamic laws.
We show that the conventional first law of thermodynamics is retrieved once the \textit{thermodynamic volume}, which is in general different from the geometric volume, is appropriately identified. 
For demonstrations of our approach, we consider familiar examples of self-gravitating gas in GR, where the connection to previous studies becomes clear. We also discuss 2-2-holes in quadratic gravity, a novel example of black hole mimickers that features super-Planckian curvatures in the interior. These objects exhibit universal high curvature effects in thermodynamics, which happen to be conveniently encoded in the thermodynamic volume. Interesting connections to black hole thermodynamics also emerge when the physical mass is treated as the total internal energy.  
\end{abstract}

%\maketitle
%%%%%%%%%%%%%%%%%%%%%%%%%%%%%%%%%
%To change the margin just for this page put the \newgeometry thing below right above the title and put the \restoregeometry in the end of the page-you gotta adjust that last part by hand. This is for that identified part arXiv put in the left hand side
%%%%%%%%%%%%%%%%%%%%%%%%%%%%%%%%%%%%%%%%%%%%%%%
%\newgeometry{left=1.8cm,right=1.5cm,bottom=2.5cm,top=2.8cm} 
\newpage
{
  \hypersetup{linkcolor=black}
  \tableofcontents
}
%\tableofcontents
%\vspace{0.5cm}
%\hrulefill
%\vspace{1cm}

\section{Introduction\label{sec:intro}}

The key ingredient behind black hole thermodynamics is the event horizon~\cite{Bardeen:1973gs,Bekenstein:1973ur,Bekenstein:1974ax,Hawking:1975vcx,Hawking:1976de,Wald:1999vt}, which leads to the information loss problem (see Ref.~\cite{Almheiri:2020cfm} for a review). Yet, current observations of astrophysical black holes in the electromagnetic and gravitational wave windows only confirm the GR predictions at the order of the horizon size with no direct implications regarding physics immediately outside the horizon.
This motivates a closer investigation of the possibility of horizonless ultracompact objects being the endpoint of gravitational collapse, which provides a resolution for the information loss problem. A variety of theoretical candidates have been proposed, and their potential observational consequences have been studied (see Ref.~\cite{Cardoso:2019rvt} for a review).  Among all, the global thermodynamics of ultracompact objects, analogous to black hole thermodynamics, has received relatively little attention.
Without the event horizon, the nontrivial contribution from background spacetime is absent~\cite{Gibbons:1976ue}, and the focus is rather on the matter source contribution and how conventional thermodynamics is influenced by curvature effects.

The statistical thermodynamics of self-gravitating systems in GR has long been explored. It was shown that the maximum entropy principle of statistical mechanics could be used to derive the basic equations describing a static and spherically symmetric self-gravitating gas in GR~\cite{Tolman:1930zza,PhysRev.36.1791} (see Refs.~\cite{Chavanis:2019ouz,Chavanis:2019mkk,Green:2013ica} for  recent reviews). Explicit examples include the self-gravitating black-body radiation~\cite{Sorkin:1981wd,Chavanis:2007kn} and fermion gas~\cite{Bilic:1999zk}, corresponding to the equilibrium solutions for photon stars and neutron stars, respectively. As a result, the total internal energy for the system is identified as the physical mass of the object, and an intimate relation between thermodynamic stability and dynamical stability is revealed. This method, nonetheless, has some limitations~\cite{Chavanis:2019ouz}. 
Since the Einstein equations are implicitly assumed in the derivation, this procedure may not be attainable for theories of modified gravity, given the more complicated structure of field equations. Also, not all information encoded in the field equations can be derived from the maximum entropy principle. 
%The Einstein equations are implicitly assumed in the derivation and not all information of the field equations can be obtained from the maximum entropy principle;  more generally, considering theories of modified gravity, this procedure may not be attainable, given the more complicated structure of field equations. 

In this paper, we study the thermodynamics of self-gravitating systems from a different perspective. Instead of deriving the profile of local thermodynamic quantities from the global ones by using a subset of the gravitational field equations, we focus on the curved spacetime effects on global thermodynamic quantities for the matter source in a general theory of gravity (GR and beyond).~This is particularly relevant to horizonless ultracompact objects where the metric resembles black hole spacetime closely from the exterior, while the interior may feature high curvatures.

In this picture, without the explicit input of field equations (and hence for a general metric), we define global extensive variables for the matter source directly from the local ones, and intensive variables by the matter properties measured at spatial infinity.
The total internal energy $U$, which is defined by properly integrating the energy density  of the matter source over the relevant volume, is then used
%, instead of the physical mass $M$, 
to  establish a well-defined thermodynamic system.
 The first law of thermodynamics for global variables applies once a so-called \textit{thermodynamic volume} $V_{th}$ is appropriately identified.\footnote{In the extended black hole thermodynamics~\cite{Kastor:2009wy,Cvetic:2010jb} (see~\cite{Kubiznak:2016qmn} for a review), which concerns spacetimes with nonzero cosmological constant $\Lambda$, the pressure $P$ is identified with the cosmological constant,  the mass of the black hole is interpreted as enthalpy $H$, and the thermodynamic volume is then given as $V_{th}=(\partial H/\partial P)_S$. This definition of thermodynamic volume is apparently irrelevant to our definition of $V_{th}$ in this work.}
For instance, for photon gas with zero chemical potential, the first law takes the conventional form:  $dU=T_{\infty}dS-p_{\infty}dV_{th}$. The thermodynamic volume $V_{th}=\int_0^R dr\sqrt{g_{rr}(r)/g^3_{tt}(r)}d^3r$, which is larger than the ordinary geometric one $V_{geo}=\int_0^R dr\sqrt{g_{rr}(r)}d^3r$ in general.  
This provides a generic description of the thermodynamic behavior of the matter source for a gravitational 
object without an event horizon. 
Compared with the first law involving the physical mass, we find $dM-dU=p_{\infty}(dV_{th}-dV_{geo}g_{tt}^{-3/2}(R))$ for the self-gravitating photon gas in GR, where the difference of $V_{th}$ and $V_{geo}$ gives rise to the difference of two internal energies. 
Furthermore, when we consider a novel example of horizonless ultracompact objects, 2-2-holes, in quadratic gravity as sourced by photon gas,  the geometric volume term becomes negligible and the difference $dM-dU$ comes mostly from the $p_{\infty} dV_{th}$ term. This yields $dM=T_\infty dS$, indicating that black hole thermodynamics can manifest in other previously unexplored scenarios without the need to delve into quantum details. The close resemblance between the exterior metric of horizonless ultracompact objects and black holes, coupled with their ability to satisfy black hole thermodynamics, provides a compelling response to potential objections to the notion of horizonless configurations in the context of the information paradox.
%The thermodynamic volume contribution turns out to be crucial for understanding  their difference. 

The rest of the paper is organized as follows. In Sec.~\ref{sec:statmech}, we present the generic description of the thermodynamic behavior of the matter sources in curved spacetime. The generalization of the first law for different kinds of matter and the general rule of identifying the thermodynamic volume are also given.
%we explore the generic laws for thermodynamic variables. 
In Sec.~\ref{sec:compactobj}, we discuss concrete examples of horizonless compact objects with vacuum spacetime at infinity in different theories of gravity, including self-gravitating photon gas and Fermi gas in GR as well as 2-2-holes with various sources. We summarize in Sec.~\ref{sec:summary}. We adopt the convention $c=\hbar=k_B=1$ throughout this work unless stated otherwise. 

%%%%%%%%%%%%%%%%%%%%%%%%%%%%%%%%%%%%%%%%%%%%%%%%%%%%%%%%%%%%%%%%%%%
\section{Thermodynamics in curved spacetime and thermodynamic volume}
\label{sec:statmech}
In this section, we revisit thermodynamics and statistical mechanics for self-gravitating systems in curved spacetime. For simplicity, we restrict to static, asymptotically flat,  and spherically symmetric spacetimes, for which the line element is generically expressed as 
\begin{eqnarray}
\label{metric}
ds^2= -B(r)\; dt^2+A(r) \;dr^2+r^2 d\theta^2+r^2 \sin^2\theta d\phi^2\;,
\end{eqnarray}
and the metric functions $A(r)$ and $B(r)$ are determined through the field equations imposed by the corresponding theory of gravity. We treat the matter source as a perfect fluid, whose energy-momentum tensor is given as
%\hbar
\begin{eqnarray}
\label{stresstensor}
T_{\mu\nu} =pg_{\mu\nu}+(\rho+p)u_{\mu}u_{\nu}
\end{eqnarray}
 where $\rho$ and $p$ denote the proper energy density and the isotropic pressure, and $u^{\mu}$ is the four-velocity, normalized as $u^{\mu}u_{\mu}=-1$.  In the rest frame of the fluid,  the tensor takes the familiar form $T^{\mu}_{\;\;\nu}=\mathrm{diag} (-\rho,  p,  p,  p)$.

In the following, we first review the well-known properties of local variables in Sec.~\ref{sec:local}. Then, we turn to global thermodynamical variables in Sec.~\ref{sec:global} by assuming the additivity of the extensive variables. As a result, we find a local-global thermodynamic correspondence for a generic curved spacetime. We argue that the internal energy $U$ for the matter source (instead of the physical mass $M$) can be used to establish a well-defined conventional thermodynamic system as long as one can identify an appropriate form of thermodynamic volume. In Sec.~\ref{sec:CIideal} and Sec.~\ref{sec:QMideal}, we illustrate the idea by considering examples of non-interacting gas described by canonical and grand canonical ensembles. The additivity of global thermodynamic potentials can be verified by derivations from the global partition function. The specific form of thermodynamic volume is also displayed.

%%%%%%%%%%%%%%%%%%%%%%%%%%%%%%%%%%%%%%%%%%%%%%%%%%%%%%%%%%%%%%%%%%% \section{Statistical mechanics in curved spacetime\label{sec:statmech}}
%\subsubsection{Canonical ensemble}
\subsection{Properties of local variables}
\label{sec:local}

In curved spacetime, a fluid element sufficiently small can be described by local thermodynamic variables in the local rest frame of the fluid element~\cite{Misner:1973prb}. According to the equivalence principle, these local thermodynamic variables shall respect the fundamental laws of thermodynamics. Here, we take two basic relations as the starting point. One is  the local Euler equation for thermodynamics~\cite{DeGroot:1980dk}, relating the proper quantities in the local frame, 
\begin{eqnarray}
\label{Gibbs-Duhem}
%\bar{s}=\frac{\rho+p}{nT}-\frac{\mu}{T}\;,
s=\frac{\rho+p-\mu\, n}{T}\;,
\end{eqnarray}
where $s$ is the entropy density, $n$ is the number density, $\mu$ is the chemical potential and $T$ is the temperature. 
The other relation comes from the local version of the first law (together with the second law) of thermodynamics~\cite{Weinberg:1972kfs-2}
\begin{eqnarray}
\label{mastereq}
%d\rho=\frac{\rho+p}{n} dn+n T d\bar{s}
d\rho=\frac{\rho+p-s\,T}{n} dn+T d s= \mu\, d n+ T\, d s\,,
\end{eqnarray}
where the energy density $\rho$ is treated as a function of number density $n$ and entropy density $s$.

For the canonical ensemble, the local thermodynamic potential is the Helmholtz free energy density $f$. It is clear from the fundamental equation that $f=\rho-s\, T$ is a function of $n$ and $T$. Following Eq.~(\ref{mastereq}), we have
\begin{eqnarray}
\label{fdiff}
%df=\frac{p}{n^2} dn-\bar{s} dT\;,
df=\mu\, dn-s \,dT\;,
\end{eqnarray}
from which the chemical potential and entropy density can directly be found as 
\begin{eqnarray}
\label{df1}
%p= n^2 \left(\frac{\partial f}{\partial n}\right)_T\label{pgen}\;,\quad
%\bar{s}= -\left(\frac{\partial f}{\partial T}\right)_n\;.
\mu= \left(\frac{\partial f}{\partial n}\right)_T\;,\quad
s= -\left(\frac{\partial f}{\partial T}\right)_n\;.
\end{eqnarray}
Using Eq.~(\ref{df1}) and  the local Euler equation, given in Eq.~(\ref{Gibbs-Duhem}), together with the definition of $f$, 
one obtains the energy density $\rho$ and the pressure $p$ in terms of $f$ as
\begin{eqnarray}
\label{df2}
%\rho=-n T^2 \frac{\partial}{\partial T}\left(T^{-1} f \right)_n\label{rhogen}\;,\quad
%\mu = f+n\left(\frac{\partial f}{\partial n}\right)_T\;.   
\rho=-T^2 \frac{\partial}{\partial T}\left(T^{-1} f \right)_n\,,\quad
p = -\frac{\partial }{\partial n}\left(n\,f\right)_T\;.   
\end{eqnarray}

Similarly, for the grand canonical ensemble, we consider the grand potential density $\omega=\rho-s\,T-\mu\,n=-p$. 
As a function of $T$ and $\mu$, the total differential is
\begin{eqnarray}
\label{grandpotdiff}
d\omega=-s \,dT-n\,d\mu\;.
\end{eqnarray}
In analogy with Eqs.~(\ref{df1}) and (\ref{df2}), we obtain the derived quantities as follows: 
\begin{eqnarray}
\label{eosgrand}
n= -\left(\frac{\partial \omega}{\partial \mu}\right)_T\;,\quad
s= -\left(\frac{\partial \omega}{\partial T}\right)_\mu\,,\quad
p=-\omega\,,\quad
\rho=-T^2 \frac{\partial}{\partial T}\left(T^{-1} \omega \right)_\mu+n\,\mu\,.
\end{eqnarray}

The spacetime variation of local thermodynamic quantities is governed by the conservation law of the stress tensor,  $\nabla^{\mu}T_{\mu\nu}=0$. For a static and spherically symmetric spacetime, this yields a single constraint from the momentum conservation along the radial direction. For the metric in Eq.~(\ref{metric}) and stress tensor in Eq.~(\ref{stresstensor}), this constraint is obtained as
\begin{eqnarray}
\label{encons0}
p'+(p+\rho)\frac{B'}{2B}=0\;,
\end{eqnarray}
where $(')$ denotes derivative with respect to $r$. 

As recently emphasized in Ref.~\cite{Lima:2019brf}, the momentum conservation of the stress tensor is directly related to Tolman's law for local thermal equilibrium in curved spacetime. 
With use of the local Euler equation, given in Eq.~(\ref{Gibbs-Duhem}), Eq.~(\ref{encons0}) can be expressed as 
\begin{eqnarray}
%\frac{B'}{2B}=-\frac{1}{\rho+p}\left(-\rho'+\bar{s}' n T+\frac{n'}{n}(\rho+p)\right)-\frac{T'}{T}-\frac{nT}{\rho+p}\left(\frac{\mu}{T}\right)'\;.
\frac{B'}{2B}=-\frac{1}{\rho+p}\left(-\rho'+T\,s' +\mu\,n'\right)-\frac{T'}{T}-\frac{n\,T}{\rho+p}\left(\frac{\mu}{T}\right)'\;.
\end{eqnarray}
The term in the parenthesis vanishes due to the first law given in Eq.~(\ref{mastereq}), and therefore
\begin{eqnarray}
\label{TolmanGen0}
\frac{B'}{2B}+\frac{T'}{T}&=&-\frac{n \, T}{\rho+p}\left(\frac{\mu}{T}\right)'\;.
\end{eqnarray}
When the chemical potential is zero, as for the photon gas, Eq.~(\ref{TolmanGen0}) reduces to the commonly known Tolman's law~\cite{Tolman:1930zza,PhysRev.36.1791}: $T(r)\sqrt{B(r)}=T_{\infty}$. The constant $T_{\infty}$ is the temperature of the gas at spatial infinity or the redshifted temperature for observers at spatial infinity. More generally, the quantity $\mu/T$, the exponential of which is commonly called \textit{fugacity}, is position independent, \textit{i.e.} $(\mu/T)'=0$, as the condition of vanishing heat flow and diffusion for a system in equilibrium~\cite{Israel:1976tn}. This then leads to the generalized version of Tolman's law~\cite{RevModPhys.21.531}
\begin{eqnarray} 
\label{eq:Tolman}
 T(r)\sqrt{B(r)}=T_{\infty}\,,\quad
 \mu(r)\sqrt{B(r)}= \mu_{\infty}\,.
\end{eqnarray}
This means that the temperature and chemical potential as intensive quantities are each uniquely specified by single numbers, $T_\infty$ and $\mu_\infty$, respectively. Since we consider only equilibrium thermodynamics, we employ the condition of constant fugacity throughout this work.

%%%%%%%%%%%%%%%%%%%%%%%%%%%%%%%%%%%%%%%%%%%%%%%%%%%%%%%%%

\subsection{The global picture and the role of thermodynamic volume} 
\label{sec:global}

The global thermodynamic characteristics of a strongly gravitating system are of interest to observers far away from the gravitational potential. For a generic discussion here, we write down global extensive thermodynamic variables in terms of their local counterparts by assuming additivity~\cite{Weinberg:1972kfs}. This is expected for a variety of matter sources and will be explicitly verified for non-interacting gas  in later subsections. Thermodynamic potentials are then obtained by appropriately integrating the corresponding potential energy density such as
\begin{eqnarray}
\label{Helmholtzglobal}
U=\int_0^R \sqrt{A\,B}\, \rho  \,d^3r\;,\quad
F=\int_0^R \sqrt{A\,B}\, f  \,d^3r\;,\quad
\Omega=\int_0^R \sqrt{A\,B}\, \omega  \,d^3r\;,
%U&=&\int_0^R \sqrt{-g}\, \rho \,d^3r=\int_0^R \sqrt{A\,B}\, \rho  \,d^3r\;,\nonumber\\
%F&=&\int_0^R \sqrt{-g}\, f  \,d^3r=\int_0^R \sqrt{A\,B}\, f  \,d^3r\;,\nonumber\\
%\Omega&=&\int_0^R \sqrt{-g}\, \omega \,d^3r=\int_0^R \sqrt{A\,B}\, \omega  \,d^3r\;,
\end{eqnarray}
where the factor $\sqrt{A\,B}$ comes from the term $\sqrt{-g}$, with $g$ being determinant of the metric, and $R$ denotes the boundary of the matter distribution. $U$, $F$, and $\Omega$ are the total internal energy, Helmholtz free energy, and the grand potential, respectively. The energy densities multiplied by $\sqrt{B}$ can be interpreted as the redshifted values measured at spatial infinity. The total number of particles $N$ and the entropy $S$ are defined through the spatial integral of the temporal components of the conserved currents $J^{\mu}$ and $S^{\mu}$ (with $\sqrt{-g}$ in the integrand). Considering that the densities are given as  $n=-J^{\mu}u_{\mu}$ and  $s=-S^{\mu}u_{\mu}$, and that the four-velocity in the rest frame of fluid becomes $u^{\mu}=(1/\sqrt{B},0,0,0)$, it is obtained that
\begin{eqnarray}
\label{NSexp}
N=\int_0^R \sqrt{A}\,n  \,d^3r\,,\quad
S=\int_0^R \sqrt{A}\,s  \,d^3r\,,
\end{eqnarray}
where the factor $\sqrt{B}$ drops out from each integrand and the volume element becomes the geometric one.

Let's consider the canonical ensemble to verify the conventional thermodynamics.
Given the Helmholtz free energy  density $f=\rho-s\,T$ and the Tolman's law Eq.~(\ref{eq:Tolman}), we can first obtain the fundamental equation 
\begin{eqnarray}\label{eq:Ffund}
F=\int_0^R \sqrt{A\,B}\, (\rho-s\, T)  \,d^3r=U-T_{\infty} S\;.
\end{eqnarray}
Then, taking the total differential of the Helmholtz free energy $F$ and implementing Eqs.~(\ref{fdiff}) and (\ref{eq:Tolman}), we have
\begin{eqnarray}
dF&=&\int_0^R \sqrt{A\,B}\, (\mu\, dn-s \,dT)  \,d^3r+ (dF)_{f}\nonumber\\
%\left.dF\right|_{f}
&=&\mu_\infty \int_0^R \sqrt{A}\, \, dn  \,d^3r-\int_0^R \sqrt{A\,B}\, s \,\frac{dT_\infty}{\sqrt{B}} \,d^3r+  (dF)_{f}\nonumber\\
&=& \mu_\infty\, dN -S\, dT_\infty+ (dF)_{N,T_\infty}\,.
\end{eqnarray}
The last term $(dF)_{N,T_\infty}$ denotes the variation of the metric function or size of the system independent of $N$ and $T_\infty$. If we attribute this change to the variation of a thermodynamic volume element
\begin{eqnarray}
\label{dVth0}
dV_{th} = -p_{\infty}^{-1} \left(dF\right)_{T_{\infty}, N}\;,
\end{eqnarray}
the conventional equation can be recovered 
\begin{eqnarray}\label{eq:dF}
dF=-S dT_{\infty}-p_{\infty}dV_{th}+\mu_\infty  dN\;.
\end{eqnarray}
Thus, by considering $F$ as a function of $N$, $T_\infty$ and $V_{th}$, one gets the consistent picture with the desired  relations
\begin{eqnarray}
\label{statefunctionsF}
S=-\left(\frac{\partial F}{\partial T_{\infty}}\right)_{V_{th},N}\;,\qquad
p_{\infty}=-\left(\frac{\partial F} {\partial V_{th}}\right)_{T_{\infty},N}\;,\qquad
\mu_\infty=\left(\frac{\partial F} {\partial N}\right)_{T_{\infty},V_{th}}\,.
\end{eqnarray}
By using these relations and Eq.~(\ref{eq:Ffund}), the internal energy can be simply expressed as $U=-T_{\infty}^2\;(\partial (T_{\infty}^{-1} F)/\partial T_{\infty})_{V_{th},N}$.
Then, with Eqs.~(\ref{eq:Ffund}) and (\ref{eq:dF}), one obtains the fundamental thermodynamic relation for internal energy, 
\begin{eqnarray}\label{eq:dU}
\label{firstlaw}
dU=T_\infty dS-p_\infty dV_{th}+\mu_\infty dN\,,
\end{eqnarray}
as the manifestation of the first law of thermodynamics for global variables.

Similarly, for the grand canonical ensemble (see Appendix~\ref{sec:microcanonical} for the microcanonical ensemble), given Tolman's law in Eq.~(\ref{eq:Tolman}), the global grand potential satisfies
\begin{eqnarray}
%\Omega=\int n \omega \sqrt{AB}\;d^3r=U-T_{\infty}S-\mu_{\infty} N\;.
\Omega=\int_0^R \sqrt{A\,B}\, (\rho-s\,T-\mu\,n) \,d^3r=U-T_{\infty}S-\mu_{\infty} N\;,
\end{eqnarray}
where the term in parenthesis is the grand potential density $w=-p$, as defined above Eq.~(\ref{grandpotdiff}). 
Together with Eq.~(\ref{eq:dF}), the total differential of $\Omega$ is given as
\begin{eqnarray}
\label{eq:dOmega}
d\Omega= -S dT_{\infty}-p_{\infty}dV_{th}-N d\mu_{\infty}\;,
\end{eqnarray}
where the thermodynamic volume can be expressed in terms of the grand potential as  
\begin{eqnarray}
\label{dVth}
dV_{th} = -p_{\infty}^{-1}(d\Omega)_{T_{\infty}, \mu_{\infty}}
= \left(d \int_0^R \sqrt{A\,B}\, \frac{p}{p_{\infty}}  \,d^3r\right)_{T_{\infty}, \mu_{\infty}}\,,
\end{eqnarray} 
and consequently, the required relations are obtained as
\begin{eqnarray}
\label{statefunctionsOmega}
S=-\left(\frac{\partial \Omega}{\partial T_{\infty}}\right)_{V_{th},\mu_\infty}\;,\qquad
p_{\infty}=-\left(\frac{\partial \Omega} {\partial V_{th}}\right)_{T_{\infty},\mu_\infty}\;,\qquad
N=-\left(\frac{\partial \Omega} {\partial \mu_\infty}\right)_{T_{\infty},V_{th}}\,.
\end{eqnarray}

In short;  we have seen that global thermodynamic variables obey conventional thermodynamics in the sense that their definitions fully encode the curved spacetime effects. The key ingredient is to appropriately identify the thermodynamic volume $V_{th}$ as in Eqs.~(\ref{dVth0}) and (\ref{dVth}). Finding the explicit expression for $V_{th}$ is not in general straightforward and its attainability highly depends on the equation of state in question. This can be conveniently seen from the definition given in Eq.~(\ref{dVth}). Unlike $T(r)$, $\mu(r)$ that follow Tolman's law, the spatial variation of pressure depends on the equation of state as we will see later in our examples. 
The difficulty is in separating the spatial integral from the $T_{\infty}$ when the latter is tangled in position-dependent non-trivial functions.
If such a separation is possible, then one can have a clear expression for $V_{th}$. For instance, in the case of a massless ideal gas with the equation of state $\rho=3p$, which yields $p\propto T^4$, we have $p(r)/p_{\infty}=B^2(r)$ from the Tolman's law, and the thermodynamic volume emerges as
\begin{eqnarray}
\label{Vth0}
V_{th}= \int_0^R \sqrt{\frac{A(r)}{B^3(r)}}\; d^3r\,,
\end{eqnarray}
with $\Omega=-p_\infty V_{th}$. Apparently, $V_{th}$ differs from the geometric volume 
\begin{eqnarray}
\label{Vgeo}
V_{geo}=\int_0^R \sqrt{A} \;d^3r\,.
\end{eqnarray}
When the curved spacetime features a deep gravitational potential, i.e. $B(r)\ll 1$, $V_{th}$ will be much larger than $V_{geo}$. 
As we will show later in Sec.~\ref{sec:compactobj} by explicit examples of compact objects with back-reaction taken into account, a larger $V_{th}$ is responsible for the difference between the physical mass $M$ and internal energy $U$. Thus, in comparison to previous studies where $M$ is interpreted as the internal energy, we keep $U$ unchanged but replace $V_{geo}$ by $V_{th}$, where the gravitational field contribution is more conveniently encoded.
%This explains the difference between our formalism for curved spacetime thermodynamics with those in the literature for self-gravitating systems in GR. In the former case, $V_{th}$ is interpreted as a generalization of the flat space volume $V$ and is related to $U$ in the usual way, while for the latter $M$ is interpreted as the internal energy. 

%%%%%%%%%%%%%%%%%%%%%%%%%%%%%%%%%%%%%%%%%%%%%%%%%%%%%%%%%%%%%%%%%%%
\subsection{Semi-classical ideal gas}
\label{sec:CIideal}

As the first example, we consider a box of a semi-classical ideal gas in thermal equilibrium. 
The local thermodynamic variables can be derived from the Boltzmann distribution in the local rest frame. For later discussion, we display expressions~\cite{DeGroot:1980dk} for the number density, energy density, pressure, and entropy density,
\begin{eqnarray}
\label{eos1}
n&=&\frac{e^{\mu/T}}{(2\pi)^3}\int e^{-E/T} d^{3}\mathrm{p}=\frac{e^{\mu/T}m^2}{2\pi^2} T K_2 (b)\,,\label{nT}\\
\rho&=&n T \left( 3+b \frac{K_1(b)}{K_2 (b)}\right)\;,\label{rhonT}\\
p&=&nT\;,\label{pnT}\\
s&=& n\left(4+b\frac{K_1 (b)}{K_2 (b)}-\frac{\mu}{T}\right)\,,\label{snT}
\end{eqnarray}
where $E=\sqrt{\mathbf{p}^2+m^2}$ is the locally measured proper energy of a particle, $K_1$, $K_2$ are the modified Bessel functions, and $b\equiv  m/T$. 
%(see also Eq.~(\ref{eosmicro}) for the corresponding equations in the massless limit). 

To complete our previous derivation of consistency between local and global pictures, we first verify the additivity of the global Helmholtz free energy in Eq.~(\ref{Helmholtzglobal}). The starting point is the fundamental equation in the canonical ensemble, namely,
\begin{eqnarray}
\label{GlobalHelmholtz}
F=-T_{\infty} \ln Z_{N}(T_\infty)\;,
\end{eqnarray}
where $Z_N$ is the N-particle global partition function. 
As usual, we evaluate the one-particle partition function $Z_1$ first. Taking the energy eigenstates of the Hamiltonian $\hat{H}$, we have
\begin{eqnarray}
Z_1(T_\infty)=\operatorname{Tr}[e^{-\hat{H}/T_\infty}]
=\sum_\mathbf{l}  e^{-E_{\mathbf{l},\infty}/T_\infty}
=\frac{1}{(2\pi)^3}\int e^{-E_\infty/T_\infty} g(E_\infty) dE_\infty\,,
\end{eqnarray} 
where $\mathbf{l}=(l_x, l_y, l_z)$ labels the momentum eigenstates in the box, and the summation over $\mathbf{l}$ is approximated by an integral for the box sufficiently large. 
$E_\infty=-\xi^\mu p_\mu$ is the conserved energy, where $\xi^\mu=(1,\mathbf{0})$ is the timelike killing vector of the static spacetime and $p^\mu$ is the particle's four-momentum. 
The density of states available to one-particle $g(E_\infty)$ for a given energy $E_\infty$ can be obtained  by $g(E_\infty)=dP(E_\infty)/dE_\infty$, with the invariant phase space volume~\cite{Padmanabhan:1989qn, Kolekar:2010py} 
\begin{eqnarray}
\label{phasespace}
P(E_\infty)=\int d^3r \,d^3\mathrm{p}\, \Theta(E_\infty+\xi^\mu p_\mu)
=\frac{4}{3}\pi \int (E_\infty^2/B-m^2)^{3/2} \sqrt{A}\, d^3r \,,
\end{eqnarray}
where $\Theta$ is the Heaviside step function accounting for the volume of momentum space with energy less than or equal to $E_\infty$.
The one-particle partition function is then
\begin{eqnarray}\label{eq:Z1}
Z_1(T_\infty)=\frac{1}{(2\pi)^3}\int e^{-E_\infty/T_\infty}\, \frac{dP(E_\infty)}{d E_\infty}\, d E_\infty=\frac{1}{(2\pi)^3}\int e^{-E/T}  \sqrt{A}\; d^3r\; d^3 \mathrm{p}
=N e^{-\mu_\infty/T_\infty}\,,
\end{eqnarray}
where Eqs.~(\ref{NSexp}) and (\ref{nT}) are used in the last step to relate $Z_1$ and $N$. Note that  $E=E_\infty/\sqrt{B}$ is the proper energy and so $E/T=E_\infty/T_\infty$. The physics behind the metric dependence for $E$ and $T$ is different. $E=E_\infty/\sqrt{B}$ is gravitational redshift due to the change of reference frames, while $T=T_\infty/\sqrt{B}$ is the Tolman's law due to thermodynamic equilibrium~\cite{Kolekar:2010py,Santiago:2018kds}.

As in the case of flat spacetime, the N-particle global partition function, $Z_N$, in the semi-classical limit is related to the one-particle partition function by $Z_N\approx Z_1^N/N!\approx (Z_1e/N)^N$, where the Stirling approximation~\cite{Stirling} is used. Thus, with Eqs.~(\ref{pnT}), (\ref{GlobalHelmholtz}) and (\ref{eq:Z1}), the global Helmholtz free energy is
\begin{eqnarray}
F&\approx&-T_\infty N\left(1+\ln\frac{Z_1}{N} \right)=N(\mu_\infty-T_\infty)\nonumber\\
&=&\int_0^R\sqrt{AB}\,(n\,\mu-p)\,d^3r=\int_0^R\sqrt{AB}\,f\,d^3r\,,
\end{eqnarray}
where the local Euler equation, given in Eq.~(\ref{Gibbs-Duhem}), and $f=\rho-s\,T$ are used in the last step. This supports the earlier definition Eq.~(\ref{Helmholtzglobal}) based on additivity.

Next, let us examine the explicit expressions for global variables. Considering the massless particle case first, 
the one-particle partition function is given through Eq.~(\ref{eq:Z1}) as 
\begin{eqnarray}
Z_1(T_\infty)=\frac{1}{\pi^2}T_\infty^3\int_0^R \sqrt\frac{A(r)}{B^3(r)}\,d^3r
\equiv \frac{1}{\pi^2}T_\infty^3 V_{th}\,.
\end{eqnarray}
In the last step, the thermodynamic volume is directly identified as that in Eq.~(\ref{Vth0}) since the $T_\infty$ dependence is fully separable from the spatial integral. 
%
%\begin{eqnarray}
%\label{Vth0}
%V_{th}= \int_0^R \sqrt{\frac{A(r)}{B^3(r)}} \;d^3r\,.
%\end{eqnarray}
%
This is consistent with the general definition of $dV_{th}$ in Eq.~(\ref{dVth0}) from the global Helmholtz free energy $F$. 
To see this, we first write down $F$ as a function of $N$, $T_\infty$ and $V_{th}$,
\begin{eqnarray}
F=-T_{\infty} \ln Z_{N}(T_\infty)
=-T_{\infty} N \left(1+\ln \frac{T_{\infty}^3 V_{th}}{\pi^2 N}\right)\,.
\end{eqnarray}
Eq.~(\ref{dVth0}) implies that the intensive quantity $p_\infty$ is the conjugate to $V_{th}$ from Eq.~(\ref{statefunctionsF}),
\begin{eqnarray}\label{eq:pinfty1}
p_\infty=-\left(\frac{\partial F} {\partial V_{th}}\right)_{T_{\infty},N}=\frac{N T_\infty }{V_{th}}\,.
\end{eqnarray}
On the other hand, in the massless limit, the local pressure $p\propto T^4$ from Eq.~(\ref{nT}--\ref{snT}) and satisfies $p(r) B^2(r)=p_{\infty}$  from Tolman's law Eq.~(\ref{eq:Tolman}). Together with Eq.~(\ref{pnT}), we can find 
\begin{eqnarray}
p_\infty V_{th}= \int_0^R \sqrt{\frac{A(r)}{B^3(r)}} B(r)^2\,n(r)\, T(r)\;d^3r
=T_\infty N\,.
\end{eqnarray}
This agrees with the derivative definition in Eq.~(\ref{eq:pinfty1}), and so it validates the $V_{th}$  definition in Eq.~(\ref{Vth0}). This shows that the thermodynamic system mimics the one in flat spacetime with the extensive global variables properly encoding the curvature effects and the intensive quantities given by the ideal gas properties at spatial infinity.

Then from Eq.~(\ref{statefunctionsF}), we can find the total entropy and chemical potential from $F$ as 
\begin{eqnarray}
S&=&-\left(\frac{\partial F}{\partial T_{\infty}}\right)_{V_{th},N}
%=\left(1-\beta_{\infty}\frac{\partial}{\partial\beta_{\infty}}\right)_{I,N}\ln Z_N
=N\left(4+\ln\left[\frac{T_{\infty}^3 V_{th}}{\pi^2 N}\right]\right)\label{Smasslessglobal}\,,\\
%p_{\infty}&=&-\left(\frac{\partial F} {\partial V_{th}}\right)_{T_{\infty},N}=\frac{N T_{\infty}}{V_{th}}\label{pmasslessglobal}\\
\mu_\infty&=&\left(\frac{\partial F} {\partial N}\right)_{T_{\infty},V_{th}}=-T_\infty \ln\left[\frac{T_{\infty}^3 V_{th}}{\pi^2 N}\right]\,.\label{mumasslessglobal}
\end{eqnarray}
As expected $S$, derived in this way, agrees with the definition in Eq.~(\ref{NSexp}) with the local variable $s=n(4-\mu/T)$ from Eq.~(\ref{snT}). The internal energy is obtained as 
\begin{eqnarray}
\label{energyglobalmassless}
U=F+T_\infty S=3 N T_{\infty}\,,
%$U&=&\left(\frac{\partial\; \beta_{\infty}F}{\partial\beta_{\infty}}\right)_{V_{th},N}=3 N T_{\infty}\nonumber\\$
\end{eqnarray}
which agrees with the local definition in Eq.~(\ref{Helmholtzglobal}) as well, with $\rho=3nT$ from Eq.~(\ref{rhonT}). The first law of thermodynamics, given in Eq.~(\ref{eq:dU}), follows accordingly.

For the massive particle case, the one-particle partition function is 
\begin{eqnarray}
\label{partitionRG0}
Z_1=\frac{m^2T_{\infty}}{2\pi^2}\int d^3r\sqrt{\frac{A}{B}}\;K_2 (m\sqrt{B}/T_{\infty})\,,
\end{eqnarray}
and the global Helmholtz free energy $F$ is given as
\begin{eqnarray}
F=-T_{\infty} \ln Z_{N}(T_\infty)
=-T_{\infty} N \left(1+\ln\left[ \frac{m^2 T_{\infty}}{2\pi^2 N}\int d^3r \sqrt{\frac{A}{B}}K_2(m\sqrt{B}/T_{\infty})\right]\right)\,.
\end{eqnarray}
In contrast to the massless case, the thermodynamic volume $V_{th}$ cannot be simply identified as the spatial integral part in the partition function due to the $T_\infty$ dependence in the Bessel function $K_2(m\sqrt{B}/T_\infty)$. Instead, we determine its differential form by evaluating the total derivative of $F$ as given in Eq.~(\ref{dVth0}). By using the pressure at infinity $p_{\infty}$, found from Eqs.~(\ref{nT}) and (\ref{pnT}) as
\begin{eqnarray}
\label{pRL}
p_{\infty}=\frac{m^2 e^{\mu_{\infty}/T_{\infty}}}{2\pi^2} T_{\infty}^2\; K_2 (m/T_{\infty})\;,
\end{eqnarray}
we obtain the differential form of thermodynamic volume 
\begin{eqnarray}
\label{dVthmassive}
dV_{th}=-p_{\infty}^{-1} (dF)_{N,T_{\infty}}=\frac{1}{K_2(m/T_\infty)}d\left(\int d^3r \sqrt{\frac{A}{B}}K_2(m\sqrt{B}/T_{\infty}) \right)_{N,T_\infty}\;.
%=\frac{1}{b_\infty^2K_2(b_\infty)}d\left(\int d^3r \sqrt{\frac{A}{B^3}}b^2K_2(b) \right)_{T_\infty,N}\,,
\end{eqnarray}
In the massless limit,  with $K_2(b)=b^2/2$, this agrees with Eq.~(\ref{Vth0}). In general, $dV_{th}$ is sensitive to the particle mass $m$ and is quite different from the universal geometric volume. As we will show later in Sec.~\ref{sec:22hole}, the combination $p_\infty dV_{th}$ might be insensitive to the mass if ultracompact objects feature a deep gravitational potential and the spatial integral is dominated by the relativistic contribution. 

The total entropy, chemical potential, and total internal energy are derived in a similar way as 
\begin{eqnarray}
S&=&-\left(\frac{\partial F}{\partial T_{\infty}}\right)_{V_{th},N}
=N\left(4-\frac{\mu_{\infty}}{T_{\infty}}+\frac{e^{\mu_{\infty}/T_{\infty}}m^3}{2\pi^2 N}\int K_1(m\sqrt{B}/T_{\infty})\sqrt{A} \;d^3r\right)\label{Smassglobal}\,,\\
\mu_\infty&=&\left(\frac{\partial F} {\partial N}\right)_{T_{\infty},V_{th}}
=-T_\infty \ln\left[ \frac{m^2T_\infty}{2\pi^2 N}\int d^3r\sqrt{\frac{A}{B}}\;K_2(m\sqrt{B}/T_{\infty})\right]\,,\\
U&=&F+T_\infty S= 3T_{\infty} N+\frac{e^{\mu_{\infty}/T_{\infty}}m^3 T_{\infty}}{2\pi^2 }\int K_1(m\sqrt{B}/T_{\infty})\sqrt{A} \;d^3r\,,
\end{eqnarray}
which agree with the quantities obtained from the local parameters. In comparison to the massless gas,  $S$ and $U$ include additional terms that depend on the spatial integral of $K_1(b)$. This is also related to the difficulty in defining the full form of $V_{th}$ from Eq.~(\ref{dVthmassive}).

 \subsection{Quantum ideal gas} 
 \label{sec:QMideal}
 %%%%%%%%%%%%%%%%%%%%%%%%%%%%%%%%%%%%%%%%%%%%%%%%%%%%%%%%%%%%
 
When the quantum nature of the source is taken into account, the particle number can fluctuate and hence is not appropriate to be treated as a state parameter in equilibrium thermodynamics. Therefore, the grand canonical ensemble is generally used to describe quantum gases where the chemical potential $\mu$, in addition to temperature and volume, can be a state parameter that handles the change of particle number and is appropriately fixed. 
 
In similarity with the canonical ensemble, we verify the consistency between the global and local pictures in the grand canonical ensemble by demonstrating the additivity of the global grand potential $\Omega$, given in Eq.~(\ref{Helmholtzglobal}). In the global picture, as in the conventional thermodynamics~\cite{Reichl:101976}, the grand potential can be derived from the fundamental equation
\begin{eqnarray}
\Omega=-T_{\infty} \ln Z\,,  
\end{eqnarray}
where  the global partition function $Z$ is 
\begin{eqnarray}
Z=\operatorname{Tr}[e^{-(\hat{H}-\mu_{\infty}\hat{N})/T_\infty}]
=\prod_\mathbf{l}\sum_{n_\mathbf{l}}e^{-(n_\mathbf{l} E_{\mathbf{l},\infty}-\mu_\infty n_\mathbf{l} )/T_\infty}\,.
\end{eqnarray}
For a given quantum state labeled by $\mathbf{l}$, the trace is evaluated with the number of eigenstates $n_\mathbf{l}$. For Bose-Einstein and Fermi-Dirac gas, summing over $n_\mathbf{l}$ leads to 
 \begin{eqnarray}
 \label{eq:OmegaQM}
\Omega=-T_{\infty}
\begin{cases}
&g\underset{l}{\sum} \ln\left[1+e^{-(E_{l,\infty}-\mu_{\infty})/T_{\infty}}\right]\;,\quad \textrm{ Fermi-Dirac}\\
&-g\underset{l}{\sum} \ln\left[1-e^{-(E_{l,\infty}-\mu_{\infty})/T_{\infty}}\right]\;,\quad \textrm{ Bose--Einstein}\;,
\end{cases}
\label{grandpot}
 \end{eqnarray}
where $g=2\sigma+1$ denotes the multiplicity for massives particles with spin $\sigma$ and $g=2$ for massless particles. As in the case of canonical ensemble, the sum can be approximated by the integral over the density of states $P(E_\infty)$. From  Eq.~(\ref{phasespace}), one obtains that

\begin{eqnarray}
\label{eq:omega}
\Omega&=&-\int_0^R   \sqrt{AB} \,d^3r\; \dfrac{T}{(2\pi)^3}\int d^3p\;
\begin{cases}
&g \ln\left[1+e^{-(E-\mu)/T}\right]\;,\quad \textrm{ Fermi-Dirac}\\
&-g \ln\left[1-e^{-(E-\mu)/T}\right]\;,\quad \textrm{ Bose--Einstein}
\end{cases}
 \nonumber\\
&=&  \int_0^R \sqrt{AB} \;w\, d^3r
\end{eqnarray}
where $\omega=-p$ is the grand potential density. The main difference for the grand canonical ensemble here is that the connection between local and global pictures in Eq.~(\ref{eq:omega}) is established through the logarithm of the partition function, given that the sum over states appears after taking the logarithm. For canonical ensemble, on the other hand, we take the sum over states before taking the logarithm, and the additivity emerges from the overall dependence on $N$.  Note that from Eq.~(\ref{eosgrand}), by using the expression for $w$ given in Eq.~(\ref{eq:omega}), one can find the other local quantities. For instance, the common case of relativistic gas of bosons and fermions  for which we can assume for simplicity that $m, \mu \ll T$, 
%$e^{\mu/T}\rightarrow 0$~\cite{Reichl2} (since $\mu \rightarrow -\infty$ at high temperatures), 
one obtains the (average) quantities as
\begin{eqnarray}
\label{avq}
n\simeq\frac{\zeta (3)}{\pi^2}\left(g_b+\frac{3}{4}g_f \right) T^3\;,\quad s\simeq\frac{2\pi^2}{45} \left(g_b+\frac{7}{8}g_f \right) T^3\;,\quad \rho \simeq 3 p \simeq \frac{\pi^2}{30} \left(g_b+\frac{7}{8}g_f \right) T^4\;,
\end{eqnarray}
where $\zeta (3)\approx 1.2$ is the Riemann zeta function of argument $3$~\cite{Reichl:101976}, $g_b$ and $g_f$ are the total bosonic and fermionic degrees of freedom respectively, which are the generalizations of $g$ in Eq.~(\ref{eq:omega}). For pure radiation (photons), since $m=\mu=0$, the expressions in Eq. (\ref{avq}) become exact rather than approximate (with $g_b=2\;, g_f=0$), which will be our first example below.

Now we will proceed to examine some examples and identify the thermodynamic volume in each case.
\\
%%%%%%%%%%%%%%%%%%%%%%%%%%%%%%%%%%%%%%%%%%%%%%%%%%%%%%%%%%%%
%%%%%%%%%%%%%%%%%%%%%%%%%%%%%%%%%%%%%%%%%%%%%%%%%%%%%%%%%%%%%

\textbf{Photon gas:} As massless Bose-Einstein gas, the particle number of photon gas is not conserved, and so its chemical potential vanishes, i.e. $\mu=0$. The local variables in this case are given as
\begin{eqnarray}
\label{eosphoton}
\rho=3p=\frac{\pi^2}{15}T^4\;,\qquad s=\frac{4\pi^2}{45}T^3\;.
\end{eqnarray}
Plugging in the Tolman's law, given in Eq.~(\ref{eq:Tolman}), we can obtain the relations
\begin{eqnarray}
\label{eq:photonrelation}
p(r)B^2(r)=p_{\infty}\,,\quad
\rho(r)B^2(r)=\rho_{\infty}\,, \quad
s(r)B^{3/2}(r)=s_{\infty}\,,
\end{eqnarray}
where $\rho_{\infty}=3p_{\infty}=\pi^2 T^4_{\infty} /15$ and $s_{\infty}=4\pi^2 T^3_{\infty} /45$ denote the gas properties at spatial infinity. 
These relations are insensitive to the numerical coefficients in Eq.~(\ref{eosphoton}) and can be directly read from the momentum conservation law in Eq.~(\ref{encons0}), given the equation of state $\rho=3p$ and the local Euler equation, given in Eq~(\ref{Gibbs-Duhem}).

With Eqs.~(\ref{eq:omega}) and (\ref{eosphoton}), the global grand potential is obtained as
\begin{eqnarray}
\label{eq:Omega}
\Omega=-\int \sqrt{A(r)B(r)} \;p \;d^3r
=-p_\infty \int_0^R \sqrt{\frac{A(r)}{B^3(r)}} \;d^3r
\equiv-p_\infty V_{th}\,.
\end{eqnarray}
With vanishing chemical potential, $\Omega$ is a function of $T_\infty$ and $V_{th}$. The thermodynamic volume $V_{th}$ is again identified as that in Eq.~(\ref{Vth0}), consistent with the derivative relation $p_\infty=-(\partial\Omega/\partial V_{th})_{T_\infty}$ in  Eq.~(\ref{statefunctionsOmega}). It is not surprising that the thermodynamic volume of the photon gas takes the same form as in the case of the semi-classical massless gas. From Eq.~(\ref{dVth}), the expression of $V_{th}$ in  Eq.~(\ref{Vth0}) follows from the relation $p(r)B^2(r)=p_{\infty}$, as dictated by the corresponding equation of state, $\rho=3p$.
From Eqs.~(\ref{eq:Omega}) and (\ref{statefunctionsOmega}), we can obtain the total entropy and internal energy 
\begin{eqnarray}
S&=&-\left(\frac{\partial \Omega}{\partial T_{\infty}}\right)_{V_{th}}=s_{\infty} V_{th}\,,\label{eq:photonS}\\
U&=&\Omega+T_\infty S=\rho_{\infty} V_{th}\,,\label{eq:photonU} 
\end{eqnarray}
with $s_\infty$, $\rho_\infty$ given below Eq.~(\ref{eq:photonrelation}). These expressions are consistent with the expected local definitions, given in Eqs.~(\ref{Helmholtzglobal}) and (\ref{NSexp}). 
\\

%%%%%%%%%%%%%%%%%%%%%%%%%%%%%%%%%%%%%%%%%%%%%%%%%%%%%%%%%%%%
\textbf{Cold Fermi gas at $T=0$:} At zero temperature, all states of the cold Fermi gas below the cutoff Fermi energy are occupied. The local variables are then characterized by the corresponding Fermi momentum $k_F$ and the fermion mass $m$ as
\begin{eqnarray}\label{eq:FDgasprho}
p=3g \rho_c \,h_p\left(\frac{k_F}{m}\right),\quad
%=\frac{m^4}{24\pi^2}\left[x(2x^2-3)\sqrt{x^2+1}+3\sinh^{-1}x\right], \nonumber\\
\rho=3g \rho_c\, h_\rho\left(\frac{k_F}{m}\right),\quad
n=g \frac{k_F^3}{6\pi^2} \,,
%=\frac{m^4}{8\pi^2}\left[x(2x^2+1)\sqrt{x^2+1}-\sinh^{-1}x\right]
\end{eqnarray}
where $g=2$ is the multiplicity in Eq.~(\ref{eq:OmegaQM}), $\rho_c=m^4/(6\pi^2)$ is the critical density, and 
\begin{eqnarray}
h_p(x)&=&\frac{1}{8}\left[x\left(\frac{2}{3}x^2-1\right)\sqrt{x^2+1}+\sinh^{-1}x\right],\quad\nonumber\\
h_\rho(x)&=&\frac{1}{8}\left[x(2x^2+1)\sqrt{x^2+1}-\sinh^{-1}x\right].\;\;
\end{eqnarray}
With vanishing temperature, the local Euler equation now becomes  $\rho+p=n\mu$, where the chemical potential is given as $\mu=\sqrt{k_F^2+m^2}$ and Tolman's law in Eq.~(\ref{eq:Tolman}) imposes the constraint $\mu(r)\sqrt{B(r)}=\mu_\infty$.

The global grand potential is a function of the chemical potential $\mu_\infty$ and the thermodynamic volume $V_{th}$, and is given as
\begin{eqnarray}
\Omega=-\int \sqrt{A(r)B(r)} \;p \;d^3r
=-3\rho_c \int_0^R\sqrt{A(r)\,B(r)}\,h_p\left(\frac{k_F}{m}\right) \;d^3r\,.
\end{eqnarray}
Similar to the massive case for a semi-classical ideal gas, the thermodynamic volume $V_{th}$ cannot simply be identified as the spatial integral part in $\Omega$ due to the complicated $T_\infty$ dependence in $h_p(k_F/m)$. 
It is then found from the generic derivative definition Eq.~(\ref{dVth}). By using $p_\infty=3\rho_c h_p(k_{F,\infty}/m)$ from Eq.~(\ref{eq:FDgasprho}), we obtain 
\begin{eqnarray}
\label{eq:coldFDdVth}
dV_{th} &=&
d\left(\frac{1}{h_p\left(\frac{k_{F,\infty}}{m}\right)} \int d^{3} r \sqrt{A\,B}\, h_{p}\left(\frac{k_F}{m}\right)\right)_{\mu_{\infty}}\;,
%d\left(\frac{1}{\hat{h}_{\rho}\left(\frac{k_{F,\infty}}{m}\right)} \int d^{3} r \sqrt{\frac{A}{B^{3}}} \hat{h}_{p}\left(\frac{k_F}{m}\right)\right)_{\mu_{\infty}}\;,
\end{eqnarray}
where $k_F/m=\sqrt{\mu_\infty^2/(B \,m^2)-1}$ and the demanded relation, $p_{\infty}=-(\partial \Omega/\partial V_{th})_{\mu_\infty}$, is then recovered. Similarly, $dV_{th}$ is sensitive to the fermion mass, in general. 
\\

%%%%%%%%%%%%%%%%%%%%%%%%%%%%%%%%%%%%%%%%%%%%%%%%%%%%%%%%%%%%
%%%%%%%%%%%%%%%%%%%%%%%%%%%%%%%%%%%%%%%%%%%%%%%%%%%%%%%%%%%%%
\section{Examples of horizonless compact objects}
\label{sec:compactobj}

We discuss explicit examples of horizonless compact objects by starting from self-gravitating gas in GR in Sec.~\ref{sec:UCOGR}. The relation of laws presented in Sec.~\ref{sec:statmech} to those in previous studies in the literature is discussed. Then, in Sec.~\ref{sec:22hole}, we consider a novel example of horizonless ultracompact objects, 2-2-holes, in quadratic gravity. They are as compact as black holes, but feature a novel high-curvature interior. The high curvature effects turn out to make significant contributions to global thermodynamic variables, where interesting connections to black hole thermodynamics emerge.

\subsection{Compact objects in GR}
\label{sec:UCOGR}

Self-gravitating systems in GR provide us with natural examples to see the curvature effects on statistical thermodynamics. Here, we consider two concrete examples of compact objects, self-gravitating photon gas confined to a spherical box~\cite{Sorkin:1981wd} and neutron stars composed of cold Fermi gas (i.e. the Oppenheimer-Volkof model)~\cite{Oppenheimer:1939ne}. 

Considering static and spherically symmetric solutions, the Einstein field equations can be simplified by the conservation of the stress-tensor and the ToV equation, given in Eq.~(\ref{encons0}) and Eq.~(\ref{eq:ToV}) of Appendix~\ref{sec:fieldeqs}, respectively.
For a given equation of state (EoS) of the matter source, the pressure $p(r)$ and mass profiles $\mathcal{M}(r)\equiv \int_0^r 4\pi r'^2\rho(r')dr'$ can be solved simultaneously as functions of the central pressure $p_c$ at the origin. 
The solutions obey a simple scaling behavior, with the following rescaled dimensionless quantities
\begin{eqnarray}\label{eq:scalingToV0}
\tilde{p}(\tilde{r})=p(r)\,\lambda^4,\,
%\tilde{\rho}=\frac{\rho}{\lambda^4},\,
\tilde{\mathcal{M}}(\tilde{r})=\mathcal{M}(r) \,\lambda^{-2} \lp^3\,,
\tilde{r}=r\, \lambda^{-2} \lp,\,
\end{eqnarray}
uniquely determined by the field equations. A one-parameter family of solution for $p(r), \mathcal{M}(r)$ can then be obtained by scaling with an arbitrary length scale $\lambda$.   
In GR, the physical mass of the object takes a simple form
\begin{eqnarray}
M=\mathcal{M}(R)\equiv \int_0^R 4\pi r^2\rho(r)dr\,.
\end{eqnarray} 
Given that the metric determinant $\sqrt{A\,B}<1$ in the interior, the total internal energy $U$, defined in Eq.~(\ref{Helmholtzglobal}), is always smaller than the mass $M$, and its smallness reflects the contribution from the gravitational field. 

Let's first consider self-gravitating photon gas confined to a spherical box, with the EoS given in Eq.~(\ref{eosphoton}). 
Here, a finite box of radius $R$ is imposed to prevent the massless gas from spreading all the way to infinity.  
Self-gravitating photon gas confined to a spherical box is then described by a two-parameter family of solutions, and the scaling behavior in Eq.~(\ref{eq:scalingToV0}) can be used to relate solutions of different $p_c$ with $\lambda=p_c^{-1/4}$.

\begin{figure}[!h]
  \centering%
{ \includegraphics[width=5.3cm]{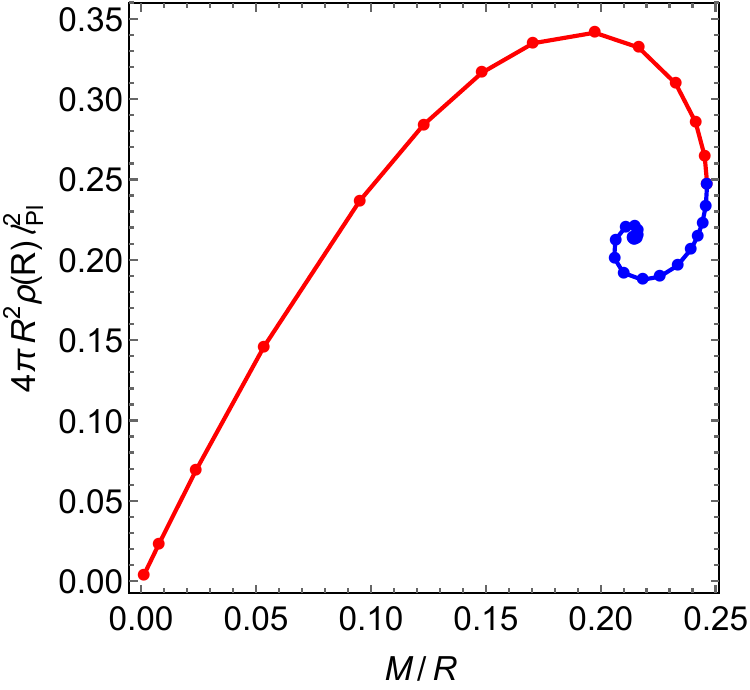}}\;
{ \includegraphics[width=5.3cm]{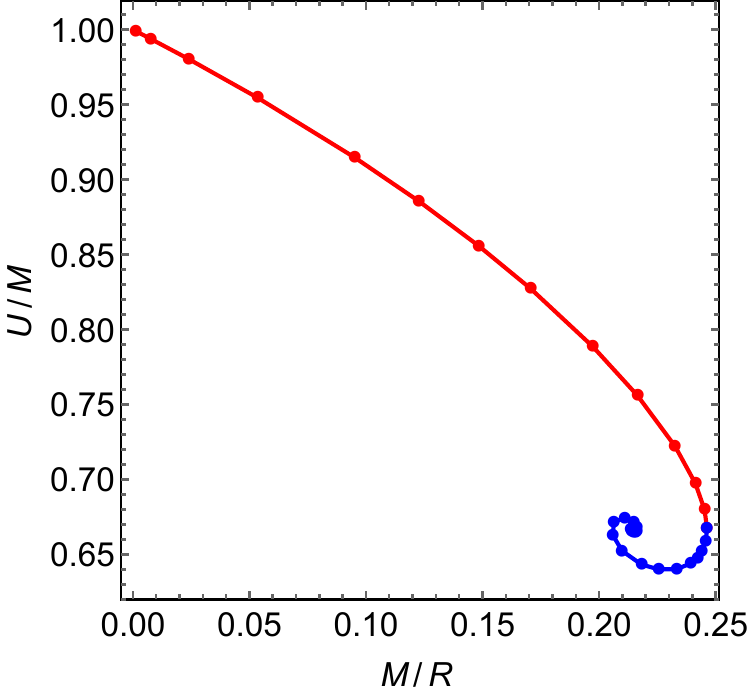}}\;
{ \includegraphics[width=5.3cm]{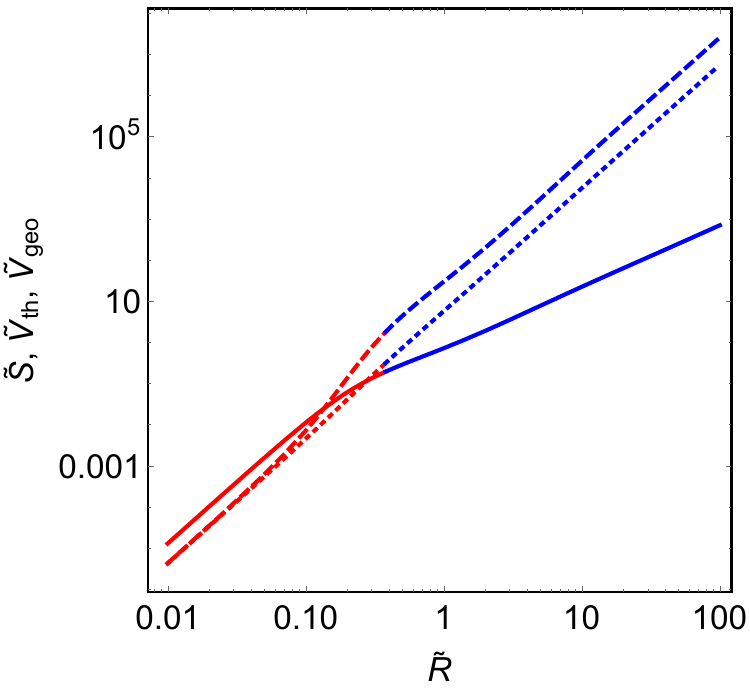}}
\caption{\label{fig:PGmodel} %Upper limits on the coupling 
Properties of self-gravitating photon gas confined to a spherical box in terms of dimensionless quantities. The red and blue denote the stable and unstable branches of the solutions. $\tilde{S}=S\, p_c^{3/4} \lp^3$, $\tilde{R}=R\, p_c^{1/2} \lp$,  $\tilde{V}=V\, p_c^{3/2} \lp^3$ are the rescaled dimensionless quantities. In the third panel, the solid, dash, and dotted lines are for $\tilde{S}$, $\tilde{V}_{th}$ (for $V_{th}$ in Eq.~(\ref{Vth0})), $\tilde{V}_{geo}$ (for $V_{geo}$ in Eq.~(\ref{Vgeo})) respectively.}
\end{figure}

Figure~\ref{fig:PGmodel} displays properties for self-gravitating photon gas. 
The global thermodynamic variables are related by $U=3\,T_\infty S/4=3\,p_\infty V_{th}$ as given in Eqs.~(\ref{eq:photonS}) and (\ref{eq:photonU}). 
When the mass-to-radius ratio $M/R$ is small, the solution describes the dilute photon gas in a box. Gravitational effects become strong with increasing $M/R$. The maximal $M/R\approx 0.25$ defines a turning point, beyond which the solutions become unstable. Along the way, the internal-energy-to-mass ratio $U/M$ decreases slowly from unity and reaches the minimum $U\approx 0.64M$ around the turning point. In the weak gravity regime, the thermodynamic volume $V_{th}$ is very close to the geometric one $V_{geo}$ and the entropy $S\propto R^3$ with $T_\infty$ roughly a constant. In the unstable branch, $V_{th}$ becomes slightly larger and $S\propto R^{3/2}$ grows slower with $R$ given $T_\infty\propto R^{-1/2}$. 

The numerical solutions can be used to examine the first law of thermodynamics. Since the solutions are described by the two parameters $p_c$ and $R$, the entropy $S$ (or the temperature $T_\infty$) and the thermodynamic volume $V_{th}$ can vary independently. The conventional first law then applies, i.e. $dU=T_\infty dS-p_\infty dV_{th}$, as we would expect from the general derivation in Sec.~\ref{sec:statmech}. 
On the other hand, it has been proved in GR~\cite{Sorkin:1981wd} that the physical mass $M$ satisfies the following first law, 
\begin{eqnarray}\label{eq:PGfirstlaw}
dM=T_\infty dS-p(R)\sqrt{B(R)} \, dV_{geo}=T_\infty dS-p_\infty\, B(R)^{-3/2} \, dV_{geo}\,,
\end{eqnarray}
where $M$ is considered as a function of $S$ and $V_{geo}$ instead. The difference between the physical mass and total internal energy is then 
\begin{eqnarray}\label{eq:dMdUdiff}
dM-dU=p_\infty \left(dV_{th}-dV_{geo}B(R)^{-3/2}\right)\,.
\end{eqnarray}
Considering the gravitational potential in the object's interior, i.e. $B(R)>B(r)$ for $r<R$, we expect $V_{th}>V_{geo}B(R)^{-3/2}$ and then $dM>dU$.
This reveals the $U$ and $M$ relation from a different perspective apart from their definitions. Interestingly, we find no discussion of such a relation between Eq.~(\ref{eq:PGfirstlaw}) and the conventional first law in the literature. 

As the second example, we consider neutron stars composed of cold Fermi gas at zero temperature, with the EoS given in Eq.~(\ref{eq:FDgasprho}). In contrast to the self-gravitating photon gas, the radius $R$ of neutron stars is determined with $p(R)=0$ and is not an independent parameter. 
Neutron stars composed of cold Fermi gas are then described by a two-parameter family of solutions, i.e. the central pressure $p_c$ and the Fermion mass $m$, and the scaling behavior in Eq.~(\ref{eq:scalingToV}) can be used to relate solutions of different mass $m$, with $\lambda=1/m$.

\begin{figure}[!h]
  \centering%
{ \includegraphics[width=5.3cm]{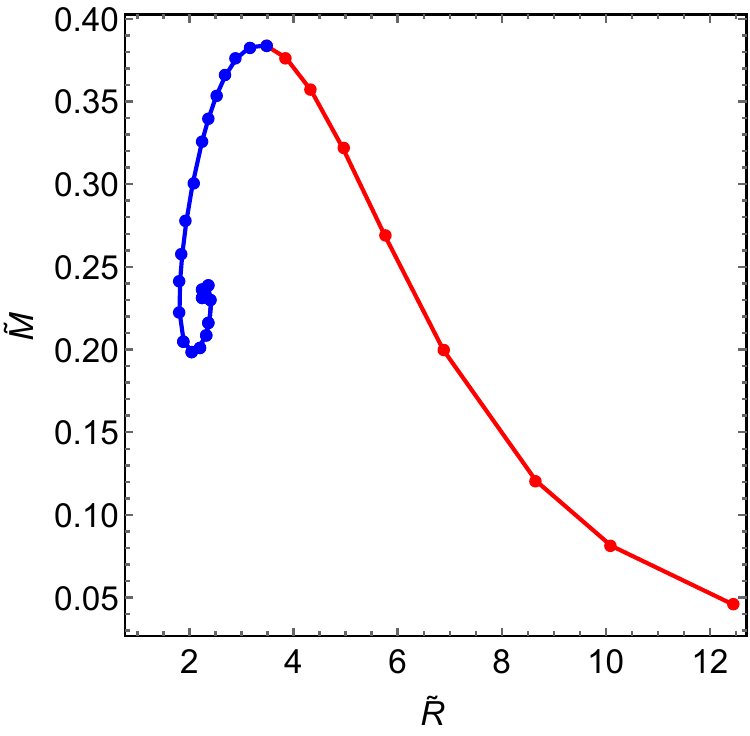}}\;
{ \includegraphics[width=5.3cm]{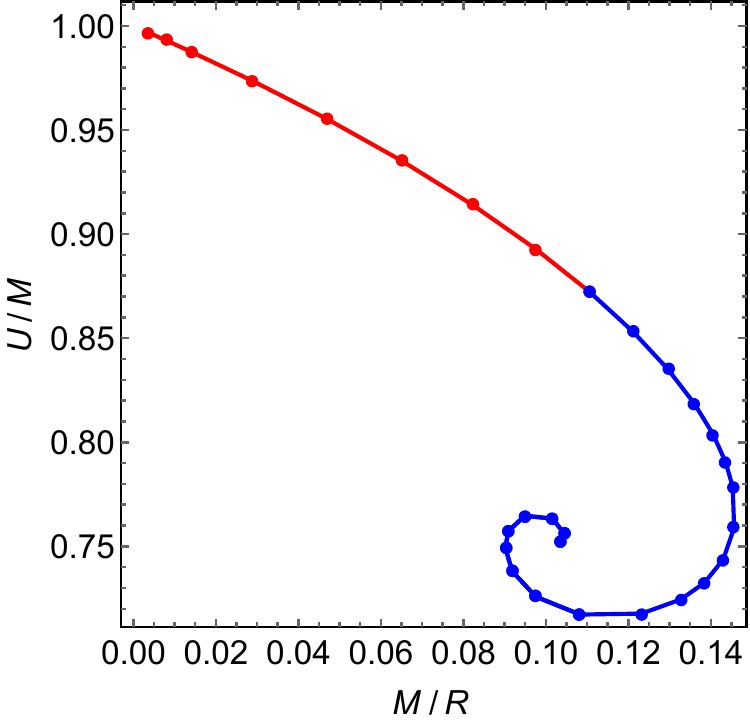}}\;
{ \includegraphics[width=5.3cm]{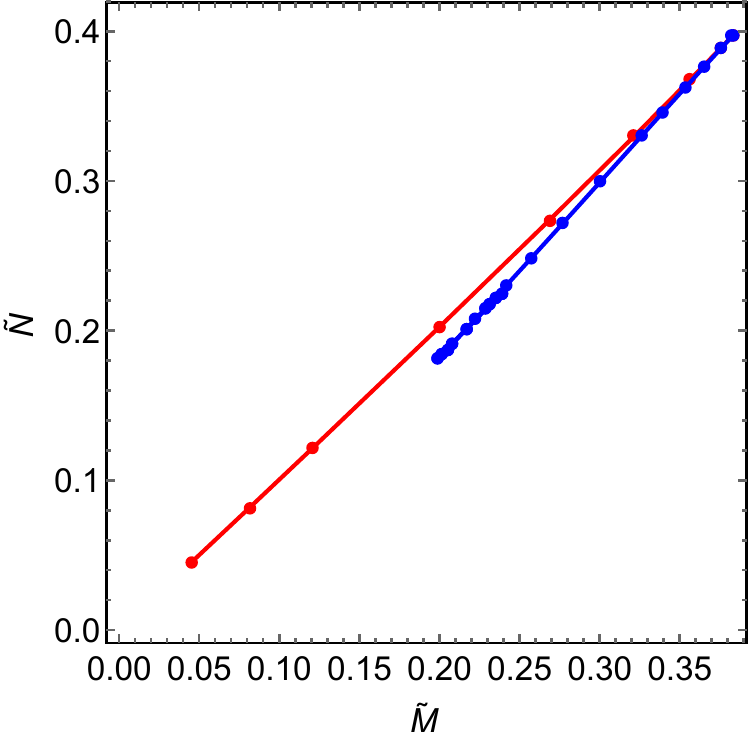}}
\caption{\label{fig:OVmodel} %Upper limits on the coupling 
Properties of neutron stars composed of the cold Fermi gas for dimensionless quantities. The red and blue denote the stable and unstable branches of the solutions. $\tilde{M}=M\, m^{2} \lp^3$, $\tilde{R}=R\, m^{2} \lp$, $\tilde{N}=N\, m^{3} \lp$ are the rescaled dimensionless quantities. }
\end{figure}

Figure~\ref{fig:OVmodel} shows properties for neutron stars composed of cold Fermi gas. Similarly, the gravitational effects become stronger with increasing central pressure. In the stable branch of solutions,  the physical mass $M$ and the total number of particles $N$ increase, and the radius $R$ decreases. In the weak gravity regime, we find $N\propto r_H/(m\,\lp^2)$, $\mu_\infty\propto r_H^0\, m$. The maximum of $M/R$ remains the turning point, but the value is slightly smaller than that of self-gravitating photon gas. The internal energy to mass ratio $U/M$ decreases from unity in a similar way and is bounded from below by $U\approx 0.72M$. 
For the first law of thermodynamics, since the solution for a given Fermion mass $m$ is described by only one parameter, the particle number $N$ and the thermodynamic volume $V_{th}$ cannot vary independently. The conventional first law Eq.~(\ref{eq:dU}) then may not be properly defined at zero temperature limit. From the numerical solutions, we find that
\begin{eqnarray}\label{eq:OVfirstlaw}
dM\approx \mu_\infty dN\,.
\end{eqnarray}  
This can be viewed as the first law built upon the physical mass $M$. 
The thermodynamic volume term in Eq.~(\ref{eq:dU}), which accounts solely for the contribution of the gravitational field, is incorporated into $dM$.
In practice, neutron stars may have a small but nonvanishing temperature. The pressure, then, would not drop to zero at a hard surface, and so a small independent $dV_{th}$ term would appear in Eq.~(\ref{eq:OVfirstlaw}).
For the unstable solutions, the global variables vary less with the central pressure and become attracted by the infinite pressure limit~\cite{Oppenheimer:1939ne}.

As a final remark, the compactness $M/R$ is sensitive to source matter properties. With exotic EoSs, it is possible to realize ultra-compact objects in GR with $R$ smaller than the photon sphere radius, i.e. $M/R\gtrsim0.33$, or even black hole mimickers with $M/R\sim 0.5$ and no horizon~\cite{Cardoso:2019rvt}. Nonetheless, no candidate makes connections between black hole thermodynamics and the conventional one as we focus on here. In the following subsection, we will discuss a candidate in a theory of modified gravity, where such a connection appears due to super-Planckian high curvature effects.

\subsection{Not quite black holes in quadratic gravity}
\label{sec:22hole}

An interesting candidate for horizonless ultracompact objects is a new type of solution, 2-2-holes~\cite{Holdom:2002xy, Holdom:2016nek}, in quadratic gravity as described by the classical action, given in Eq.~(\ref{eq:CQG}) in Appendix \ref{sec:fieldeqs}.  The existence of 2-2-holes relies on the Weyl term $C^{\mu\nu\rho\sigma}C_{\mu\nu\rho\sigma}$,  which introduces a new spin-2 mode with the Compton wavelength $\lambda_2$. 
In the presence of a compact matter source, a typical 2-2-hole with mass $M$ much larger than the minimum mass $\Mmin\sim \Mp^2 \lambda_2$ resembles a black hole closely from the exterior, while a novel interior takes over right above the would-be horizon $r_H=2M\lp^2$. The interior curvature can easily reach super-Planckian values when $\lambda_2\sim \lp$, and approaches infinity at the origin.  
In contrast to proposed ultracompact objects in GR, the 2-2-hole properties are crucially determined by the high curvature interior, while being quite insensitive to the details of EoS.  
Thus,  2-2-holes provide a tractable model to study the influence of high curvature effects on statistical thermodynamics. In the following, we consider concrete examples of 2-2-holes sourced by various simple forms of matter.   

 First, we consider thermal 2-2-holes sourced by massless particles, i.e.~photon gas\footnote{Basic thermodynamic relations for 2-2-holes sourced by photon gas were studied in Refs.~\cite{Holdom:2019ouz, Ren:2019afg}. Here, while reproducing some of their results, we discuss the corresponding implications with the generic descriptions given in this paper. We also present results for 2-2-holes sourced by cold Fermi gas.} and massless classical ideal gas, \textit{e.g.} EoS being $\rho_{\infty}=3p_{\infty}$. 
The gas profile and the metric functions can be solved from the momentum conservation of the stress-tensor Eq.~(\ref{encons0}), which yields $p(r)B(r)^2=p_\infty$ for relativistic gas, and two field equations, given in Eq.~(\ref{eq:fieldeq1}) in Appendix \ref{sec:fieldeqs}.
Fig.~\ref{fig:22hole} displays the 2-2-hole solutions with two different values of mass $M\gg \Mmin$. 
Despite the complicated form of field equations in quadratic gravity, the numerical solutions of 2-2-holes turn out to be governed by simple qualitative features. 
The exterior ($r\gtrsim r_H$), where the curvature becomes lower for increasing size of the object, obeys the Einstein field equations to a good approximation and follows the conventional scaling in Eq.~(\ref{eq:scalingToV0}). The interior $r<r_H$, on the other hand, is a high curvature regime and features a huge gravitational potential. At the leading order of high curvature expansion, it is described by a novel scaling behavior, with the following rescaled dimensionless quantities~\cite{Holdom:2016nek}
\begin{eqnarray}\label{eq:22scaling}
\tilde{p}(\tilde{r})=p(r) \lambda_2^2\,\lp^2,\;
\tilde{A}(\tilde{r})=A(r) \frac{r_H^2}{\lambda^2_2},\;
\tilde{B}(\tilde{r})=B(r) \frac{r_H^2}{\lambda^2_2}
\end{eqnarray}
uniquely determined as functions of $\tilde{r}=r/r_H$. 
Similarly, a box has to be imposed to make the total energy finite, and the box radius $R$ has to be considerably larger than $r_H$ to not ruin the simple behavior.

\begin{figure}[!h]
  \centering%
{ \includegraphics[width=7.5cm]{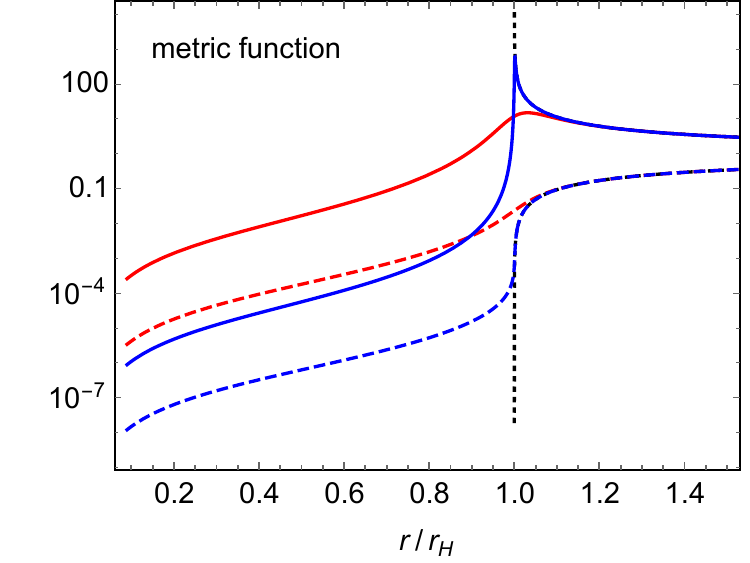}}\;
{ \includegraphics[width=7.5cm]{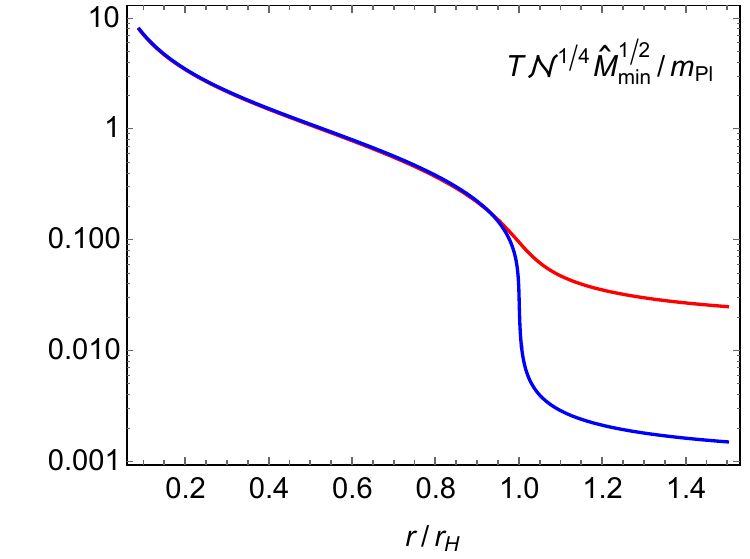}}
\caption{\label{fig:22hole} %Upper limits on the coupling 
Properties of thermal 2-2-holes sourced by relativistic gas with EoS $\rho=3p$. Left: the metric $A$ (solid) and $B$ (dashed) as functions of $r/r_H$, where $r_H$ is the would-be horizon size. Right: the thermal gas temperature $T$ as a function of $r/r_H$. The red and blue are for $r_H/\lambda_2=6,\,100$ respectively. }
\end{figure}

From the novel scaling in Eq.~(\ref{eq:22scaling}) and the numerical solutions, we find the rescaled gas pressure at infinity
\begin{eqnarray}\label{eq:pinf}
\tilde{p}_\infty=\tilde{p}(\tilde{r})\tilde{B}(\tilde{r})^2=p_\infty  \frac{\lp^2}{\lambda_2^2}r_H^4\approx 1.35\times10^{-5}\,.
\end{eqnarray} 
The thermodynamic volume $V_{th}$ in Eq.~(\ref{Vth0}) can be separated into two parts, the interior contribution $V_{th}^{(in)}$ for $r\lesssim r_H$ and the exterior contribution $V_{th}^{(ex)}$ for $r_H\lesssim r \lesssim R$. The interior contribution is governed by the novel scaling in Eq.~(\ref{eq:22scaling}), satisfying 
\begin{eqnarray}\label{eq:Ivol}
\tilde{V}_{th}^{(in)}= V_{th}^{(in)}\frac{\lambda_2^2}{r_H^5}
\approx 4.63\times 10^3\,.
%= \frac{r_H^5}{\lambda_2^2}\int\sqrt\frac{A(r)r_H^2/\lambda_2^2}{(B(r) r_H^2/\lambda_2^2)^3}\,4\pi \frac{r^2}{r^2_H} \, d\frac{r}{r_H}
\end{eqnarray}
%The novel scaling $V_{th}^{(in)}\propto r_H^5/\lambda_2^2$ comes directly from the high curvature effects. 
In comparison to the flat spacetime scaling $V_{th}^{(in)}\sim r_H^3$, the high curvature effects lead to a large enhancement of order $(r_H/\lambda_2)^2$. The gain could be enormous given that $\lambda_2$ can be as small as $\lp$. The exterior contribution encodes the box radius $R$ dependence. Since the gas temperature drops drastically at $r>r_H$ as shown in Fig.~\ref{fig:22hole}, we expect a trivial dependence $V_{th}^{(ex)}\propto R^3$ from the cold and dilute gas outside as in the case of GR. 

Thus, for macroscopic 2-2-holes with $R$ not significantly larger than $r_H$, we expect $V_{th}$ to be dominated by the interior contribution. This is also true for the total internal energy and entropy of the gas, which are proportional to $V_{th}$.
With Eqs.~(\ref{eq:photonU}), (\ref{eq:pinf}), and (\ref{eq:Ivol}), the total internal energy is
\begin{eqnarray}\label{eq:UM22}
U\approx U^{(in)}= 3 \, p_\infty\, V_{th}^{(in)}\approx \frac{3}{8} M\,.
\end{eqnarray}
Since the compactness of 2-2-holes is nearly maximal, it is not surprising that its internal energy-to-mass ratio is far smaller than that for compact objects in GR. 
The same ratio has also been obtained in the brick wall model~\cite{tHooft:1984kcu}, but there, the back-reaction has not been taken into account. 
Using Eqs.~(\ref{eosphoton}), (\ref{eq:photonS}), (\ref{eq:pinf}), and (\ref{eq:Ivol}), we can obtain the temperature at spatial infinity and the entropy  
\begin{eqnarray}\label{eq:TS1}  
%T_{\infty}&=&\frac{M}{8N}
T_{\infty}\approx 1.4\,\hMmin^{1/2}\,T_\textrm{BH}\,,\quad
S\approx S^{(in)}\approx 0.71\,\hMmin^{-1/2}\,S_\textrm{BH}\,\,,
\end{eqnarray}
where $\hMmin=\Mmin/\Mp=0.63\lambda_2/\lp$, the Hawking temperature $T_\textrm{BH}=\Mp^2/8\pi M$ and the Bekenstein-Hawking entropy  $S_\textrm{BH}=\pi\, r_H^2/\lp^2$.\footnote{Throughout the paper, we consider systems in local thermal equilibrium, thus assumed that the flat spacetime equilibrium thermodynamics is valid locally due to the equivalence principle. This assumption may not apply if the local curvature is much larger than the local temperature, namely, if the thermal wavelength is much larger than the typical radius set by curvature. For thermal 2-2-holes, we can use the Weyl curvature squared $C_{\mu\nu\rho\sigma}C^{\mu\nu\rho\sigma}$ to estimate the curvature scale (instead of the Ricci scalar $R$ which is usually considerably suppressed). From the leading order of high curvature expansion, we find that $(C_{\mu\nu\rho\sigma}(r)C^{\mu\nu\rho\sigma}(r))^{1/4}\approx 1.8 (r_H/\lp)/r$ and $T(r)\approx 0.9(r_H/\lp)/r$ are on the same order of magnitude. }
As a result of high curvature effects, a 2-2-hole sourced by relativistic particles exhibits Hawking-like temperature and the entropy area law~\cite{Holdom:2016nek}. 
Given that $S=s_\infty V_{th}$ in Eq.~(\ref{eq:photonS}), we can see that the enormous amount of microscopic entropy is achieved here mainly through the huge value of $V_{th}^{(in)}$ given in Eq.~(\ref{eq:Ivol}). This is usually considered difficult for self-gravitating objects with no horizon.
The numerical values of $T_{\infty}$ and $S$ differ from the black hole counterparts due to the $\Mmin$ or $\lambda_2$ dependence. 

For the first law of thermodynamics, as 2-2-holes sourced by photon gas are described by two parameters $r_H$ and $R$, the entropy $S$ and the thermodynamic volume $V_{th}$ can vary independently, and the conventional first law still applies, i.e. $dU=T_\infty dS-p_\infty dV_{th}$, as we would expect. Given the absence of analytical solutions, the variation of the physical mass $M$ requires to be checked numerically. Using $T_\infty S^{(in)}=4\,p_\infty V_{th}^{(in)}\approx M/2$, as obtained from the relations given below Eq.~(\ref{eq:photonrelation}) and in Eq.~(\ref{eq:UM22}),  with the relations $T_\infty\propto M^{-1}$ and $S^{(in)}\propto M^2$  as can be seen in Eq.~(\ref{eq:TS1}), we obtain
\begin{eqnarray}\label{eq:firstlaw220}
dM\approx T_\infty dS^{(in)}\approx T_\infty dS\,.
\end{eqnarray}
In comparison to Eq.~(\ref{eq:PGfirstlaw}), the work performed at the box radius is numerically negligible for macroscopic 2-2-holes with $V_{geo}\sim R^3\ll V_{th}^{(in)}$. The major difference between the physical mass and total internal energy is then accounted for by the thermodynamic volume contribution $p_\infty dV_{th}\approx 5dM/8$ with $dM\approx dU+p_\infty dV_{th}$.

It is instructive to make some comparisons with black hole thermodynamics. For Schwarzschild black holes, $dM=T_\infty dS$ applies without the volume term because of the vanishing thermodynamic pressure.
 For thermal 2-2-holes,  Eq.~(\ref{eq:firstlaw220}) holds at the leading order of high curvature expansion since the volume term associated with variation of the box boundary has a negligible contribution. 
This shows that certain aspects of black hole thermodynamics could be derived from conventional thermodynamics by utilizing the non-trivial structure of curved spacetime, and thus are not too mysterious.

For 2-2-holes sourced by a semiclassical ideal gas, there is the additional dependence on the conserved number of particles $N$ or the nonzero chemical potential $\mu_\infty$ given in Eq.~(\ref{mumasslessglobal}). The relation between $U$ and $M$ remains the same as in Eq.~(\ref{eq:UM22}), while the temperature and total entropy change. With Eqs.~(\ref{nT}), (\ref{pnT}), and (\ref{Smasslessglobal}) for the massless ideal gas, we obtain
%$p_\infty=T_\infty^4/\pi^2 e^{\mu_\infty/T_\infty}$, $T_\infty S=(4-\mu_\infty/T_\infty)p_\infty V_{th}$, 
\begin{eqnarray}\label{eq:TS2}
%T_{\infty}&=&\frac{M}{8N}
T_{\infty}\approx 1.7\,\hMmin^{1/2}\,e^{-\frac{1}{4}\mu_{\infty}/T_{\infty}}\,T_\textrm{BH}\,,\;
%S&=&4N\left(1-\frac{1}{N}\frac{\mu_{\infty}}{T_{\infty}}\right)
S^{(in)}\approx 0.59\,\hMmin^{-1/2}\,e^{\frac{1}{4}\mu_{\infty}/T_{\infty}}\left(1-\frac{\mu_{\infty}}{4T_{\infty}}\right)\,S_\textrm{BH}\,\,.
\end{eqnarray}
Given that $\exp(\mu_\infty/T_\infty)\ll1$  for a semiclassical ideal gas, the corresponding thermal 2-2-holes have higher temperature and lower entropy than holes sourced by photon gas of the same mass in Eq.~(\ref{eq:TS1}). As the generalization of conventional thermodynamics, it is not surprising that thermal 2-2-hole characteristics can depend on certain aspects of matter properties. 

Considering the entropy $S$ as a function of two independent variables $M$ and $N$ as given in Eq.~(\ref{Smasslessglobal}), the total differential of $S$ for the interior contribution is
\begin{eqnarray}
dS^{(in)}\approx dN^{(in)} \ln(e^{-\mu/T})+N^{(in)}\left(5 \frac{dM}{M}+3 \frac{dM}{M}\right)
\approx-\frac{\mu_\infty}{T_\infty} dN^{(in)}+\frac{1}{T_\infty}dM\,,
%&=&dN\left[4+\ln\left(\frac{I T_\infty^3}{\pi^2 N}\right)\right]+N\left(d\ln I + d(T_\infty N)^3-d\ln N^4\right)\nonumber\\
\end{eqnarray}
with $V_{th}^{(in)}\propto M^5$ and $T_\infty N^{(in)}\propto M$. This yields a generalization of Eq.~(\ref{eq:firstlaw220}) with the additional contribution from the conserved number of particles,
\begin{eqnarray}\label{eq:firstlaw221}
dM\approx T_\infty \,dS+\mu_\infty dN\,.
\end{eqnarray}
The exterior contributions are again ignored. Since black holes do not conserve charges associated with global symmetries~\cite{Arkani-Hamed:2006emk}, this has no counterpart in black hole thermodynamics. For a macroscopic 2-2-hole, the particle-number-to-mass ratio $N\Mp/M\approx \Mp/(8T_\infty)\propto M/\Mp$ is enormous, and it can increase indefinitely with the hole mass, unlike the local charge of black holes to be bounded by the extremal limit.

2-2-holes can also be sourced by semiclassical ideal gas with nonzero mass. As the temperature satisfies Tolman's law $T(r)\sqrt{B(r)}=T_\infty$, the gas becomes relativistic around the origin, while the density and pressure are suppressed by the Boltzmann factor when $T(r)$ drops below the particle mass $m$. Although the field equations, given in Eq.~(\ref{eq:fieldeq2}) in Appendix \ref{sec:fieldeqs}, are more involved for this case, the interior metric functions and gas profile remain characterized by the same scaling behavior Eq.~(\ref{eq:22scaling}) as for relativistic thermal gas,  for a given $m\sqrt{\lambda_2\lp}$. Interestingly, global thermodynamic quantities such as $T_\infty$, $S$, $N$, and $U$ are found to be quite insensitive to the particle mass due to the dominance of the relativistic contribution. Thus, the results obtained for the massless case can be used as a reliable approximation for the massive case, with the mass dependence having a negligible impact on the leading order results. This includes the intricacy of identifying the thermodynamic volume for the massive case, as discussed around Eq.~(\ref{dVthmassive}).

Another example is 2-2-holes sourced by the cold Fermi gas, which may serve as the endpoint of the gravitational collapse of neutron stars after sufficient cooling. Together with Tolman's law Eq.~(\ref{eq:Tolman}) for chemical potential, the metric functions and gas profile are solved from the field equations (Eq.~(\ref{eq:fieldeq2}) in Appendix \ref{sec:fieldeqs}) with EoS given in Eq.~(\ref{eq:FDgasprho}). The cold Fermi gas becomes relativistic around the origin due to the quantum pressure. 
As in the case of neutron stars in GR, the pressure (and the Fermi momentum $k_F$) drops to zero at some radius $R$ and defines the object's surface. The difference is that $R$ is within the would-be horizon $r_H$ for 2-2-holes, while it is considerably larger than $r_H$ for neutron stars. 

For a given $m\sqrt{\lambda_2\lp}$, the interior is again characterized by the novel scaling behavior in Eq.~(\ref{eq:22scaling}). With this scaling, the rescaled chemical potential redshifted to infinity is
\begin{eqnarray}\label{eq:muinf}
%\mu_\infty\approx 0.22 \sqrt\frac{\lambda_2}{\lp}\frac{1}{r_H}\,,
\tilde{\mu}_\infty=\mu_\infty \sqrt{\frac{\lp}{\lambda_2}}r_H\approx 0.22\,.
\end{eqnarray}
Since $R<r_H$, the total internal energy and (average) number of particles receive contributions only from the 2-2-hole interior. The internal energy to mass ratio $U/M$ remains roughly $3/8$.  The rescaled number of particles is
\begin{eqnarray}\label{eq:Inum}
\tilde{N}=N \, \frac{\lp^{3/2}\lambda_2^{1/2}}{r_H^2}\approx 1.15\,.
%N = \frac{r_H^2}{\lp^{3/2}\lambda_2^{1/2}}\int\frac{(k_F(r)\sqrt{\lambda_2\lp})^3}{3\pi^2}\sqrt{A(r)r_H^2/\lambda_2^2}\,4\pi \frac{r^2}{r^2_H} \, d\frac{r}{r_H}
%\approx 1.15\,\frac{r_H^2}{\lp^{3/2}\lambda_2^{1/2}}\,,
\end{eqnarray}
Note that the $r_H$ (or $M$) dependences of $\mu_\infty$ and $N$ are quite different from that for neutron stars, but resemble closely that for $T_\infty$ and $S$ for thermal 2-2-holes in Eq.~(\ref{eq:TS2}). 
With Eqs.~(\ref{eq:muinf}) and (\ref{eq:Inum}), it is straightforward to obtain
\begin{eqnarray}
dM\approx \mu_\infty\, dN\,,
\end{eqnarray} 
the analog of Eq.~(\ref{eq:OVfirstlaw}) for neutron stars and of Eq.~(\ref{eq:firstlaw221}) for thermal 2-2-holes. 
Again,  the thermodynamic volume term accounts for the difference $1-U/M$, and the contribution is the same as in the case of thermal 2-2-holes.  

In summary, the global thermodynamic variables of 2-2-holes exhibit interesting universal characteristics due to the high curvature interior and its novel scaling behavior. 
It is intimately related to the greatly enhanced interior thermodynamic volume that dominates the contribution for macroscopic holes. This renders the EoS dependence rather weak, as opposed to the situation in GR. 

%%%%%%%%%%%%%%%%%%%%%%%%%%%%%%%%%%%%%%%%%%%%%%%%%%%%%%%%%%%%%%%%%%%%%%
%%%%%%%%%%%%%%%%%%%%%%%%%%%%%%%%%%%%%%%%%%%%%%%%%%%%%%%%%%%%%%%%%%%%%%
\section{Summary}
\label{sec:summary}

We have studied the thermodynamics of self-gravitating systems in a new approach to better characterize horizonless ultracompact objects in a general theory of gravity. 
%in order to explore thermodynamic characteristics of horizonless ultracompact objects in a general theory of gravity. 
In Sec.~\ref{sec:statmech}, we have derived generic thermodynamic laws of global variables for the matter source, without the explicit input of gravitational field equations, from the curved spacetime generalizations of thermodynamic potentials for different statistical ensembles. Consequently, unlike the common treatment in the literature, there is no direct reference to the physical mass and the total internal energy $U$ is defined as usual in Eq.~(\ref{Helmholtzglobal}). The conventional thermodynamic laws then arise directly from the global and local correspondence, except that thermodynamic volume $V_{th}$ has to be appropriately identified. The most generic definition of $V_{th}$ can be derived from the first law of thermodynamics, as given by  Eqs.~(\ref{dVth0}) and (\ref{dVth}) for canonical and grand canonical ensembles, respectively. For illustration, we have considered examples of non-interacting gas, where the global thermodynamic potentials can be derived from the global partition functions. The explicit form of global thermodynamic variables is displayed. For relativistic gas in particular, i.e. with EoS $\rho=3p$, it is possible to identify the explicit form of $V_{th}$, as given in Eq.~(\ref{Vth0}), with the help of Tolman's law, given in Eq.~(\ref{eq:Tolman}). It is in general larger than the geometric volume $V_{geo}$, given in Eq.~(\ref{Vgeo}), due to the extra enhancement from gravitational redshift.  

Then in Sec.~\ref{sec:compactobj}, we have studied specific examples of horizonless ultracompact objects with vacuum spacetime at infinity.
%horizonless ultracompact objects by taking into account the back-reaction.
We have first considered familiar examples of self-gravitating gas in GR, including photon gas in a box and cold Fermi gas. In comparison with the previous studies, where the physical mass $M$ of the system is identified as the total internal energy, we have highlighted the difference between the forms the first law takes in these two approaches. The difference between $M$ and $U$  turns out to be directly related to the difference between thermodynamic volume $V_{th}$ and the geometric one $V_{geo}$, as given in Eq.~(\ref{eq:dMdUdiff}). For the case where the object boundary is determined by gravitational field equations, e.g. the cold Fermi gas, $V_{th}$ does not vary independently and its contribution has to be absorbed into the variation of $M$.  

Furthermore in Sec.~\ref{sec:compactobj}, we have considered another candidate for horizonless ultracompact objects, 2-2-hole in quadratic gravity, sourced by similar kinds of gases used in the previous examples. These objects are as compact as black holes and are drastically different from the proposed ultracompact objects in GR due to the novel interior with super-Planckian curvatures. A novel scaling behavior in Eq.~(\ref{eq:22scaling}) emerges for the interior solution, to be compared with the GR scaling in Eq.~(\ref{eq:scalingToV0}). As a result, the interior thermodynamic volume $V_{th}^{(in)}$ is strongly enhanced by a factor of $r_H^2/\lp^2$ for macroscopic holes in comparison to the naive estimate of order $r_H^3$. This then leads to Hawking-like temperature and the entropy area law for 2-2-holes from conventional thermodynamics as given in Eqs.~(\ref{eq:TS1}) and (\ref{eq:TS2}). Such large $V_{th}$ is also what is necessary to account for the large difference between $M$ and $U$. Because of the dominance of the high curvature effects,  the 2-2-hole thermodynamics shows universal characteristics and is much less sensitive to matter EoS than compact objects in GR. 
Note that the peculiar thermodynamic characteristics for 2-2-holes are only verified numerically due to the absence of analytical relations between $M$ and the source properties. The possibility for a more analytical derivation, as in the case of GR, requires further investigation.

 For other theories of modified gravity with nonvacuum spacetime at infinity, e.g. scalar-tensor theories, the generic descriptions for the matter source in  Sec.~\ref{sec:statmech} still apply, with the effects of the scalar charge encoded in the metric functions.  The main change to what we presented above is the additional contribution of the scalar charge to the difference of $dM$ and $dU$ in Eq.~(\ref{eq:dMdUdiff})~\cite{Lee:1974pt, Shibata:2013ssa}. 
More generally, the exact meaning of the thermodynamic volume $V_{th}$ requires a more comprehensive understanding. In this study, the volume is modified to ensure the validity of the first law of global thermodynamics. Nevertheless, it would be intriguing to explore the implications of this modification in the context of emergent gravity from thermodynamics~\cite{Jacobson:1995ab,Padmanabhan:2002sha,Cai:2005ra,Padmanabhan:2009vy,Verlinde:2010hp}. One question that arises is whether it is feasible to derive the field equations from the modified version of thermodynamic laws. Additionally, a closer examination of fuzzballs in string theory, which serve as prominent examples of horizonless ultra-compact objects, would be captivating~\cite{Lunin:2001jy,Mathur:2014nja}. These aspects warrant more comprehensive investigations, which we leave as avenues for future research.

%%%%%%%%%%%%%%%%%%%%%%%%%%%%%%%%%%%%%%%%%%%%%%%%%%%%%%%%%%%%%%%%%%%%%%

\vspace{0.1cm}
\section*{Acknowledgements} 
\vspace{-0.1cm}
U.A. thanks Altu\u{g} \"Ozpineci for useful communications. The work of U.A. was supported in part by the Institute of High Energy Physics, Chinese Academy of Sciences, under Contract No.~Y9291220K2 (until June 2022), the Chinese Academy of Sciences President's International Fellowship Initiative (PIFI) under Grant No.~2020PM0019 (until June 2022), The Scientific and Technological Research Council of T\"urkiye (T\"UB\.ITAK), B\.{I}DEB 2232-A program, under project number
121C067 (from September 2022 on). J.R. was supported by the Institute of High Energy Physics under Contract No.~Y9291220K2.  

%%%%%%%%%%%%%%%%%%%%%%%%%%%%%%%%%%%%%%%%%%%%%%%%%%%%%%%%%%%%%%%%%%%%%%%%%%%%%%%%%%%%%%%%%%%%%%%%%%%%%%%%%%%%%%%%%

\appendix

\section{Microcanonical ensemble\label{sec:microcanonical}}
%\subsection{Zero mass case}
We have focused on canonical and grand canonical ensembles throughout the paper. Here, we simply address the other commonly used framework, microcanonical ensemble, in a simple example of a massless semi-classical ideal gas. The main thermodynamic function to be used is entropy $S$, determined by counting the number of microstates for specific parameters. The state parameters in this ensemble are total internal energy $U$, the total number of particles $N$, and the thermodynamic volume $V_{th}$. 
%The last one is a reminder that we are in curved spacetime and the global parameter that plays the role of volume in thermodynamic sense, \textit{i.e.} the conjugate of global pressure $p_{\infty}$ in the fundamental equation, is the thermodynamic volume, which in general is different than the geometric volume, used in ordinary, flat-space, thermodynamics.

Before going into the example, let's first have a general discussion. The total differential of entropy $S(U,V_{th}, N)$, directly from the fundamental relation (\ref{firstlaw}), is given as
\begin{eqnarray}
dS=\frac{1}{T_{\infty}}dU+\frac{p_{\infty}}{T_{\infty}}dV_{th}-\frac{\mu_{\infty}}{T_{\infty}}dN\;,
\end{eqnarray}
where
\begin{eqnarray}
\label{statefunctionsS}
\frac{1}{T_{\infty}}=\left(\frac{\partial S}{\partial U}\right)_{V_{th},N}\;,\qquad
p_{\infty}=T_{\infty} \left(\frac{\partial S} {\partial V_{th}}\right)_{U,N}\;,\qquad
\mu_\infty=-T_{\infty} \left(\frac{\partial S} {\partial N}\right)_{U,V_{th}}\,.
\end{eqnarray}
It is clear that in order to have a consistent picture we have to define the thermodynamic volume, in analogy to the canonical (\ref{dVth0}) and grand canonical (\ref{dVth}) cases, as
\begin{eqnarray}
\label{dVth2}
dV_{th} = \frac{T_{\infty}}{p_{\infty}}(dS)_{U, N}
= \frac{T_{\infty}}{p_{\infty}}\left(d \int_0^R \sqrt{A}\, s  \,d^3r\right)_{U, N}\,.
\end{eqnarray} 
Here again, we have the issue we ran into in the other ensembles. If the spatial dependence in $s$ can be separated from the quantities measured at infinity, we can simply find the expression for $V_{th}$, otherwise it is not clear if an explicit expression can be obtained.

For ideal gas, the total entropy is obtained as
\begin{eqnarray}
\label{Smicrostates}
%S=\ln\Omega\;,
S=\ln\mathcal{N}_N(U)\;,
\end{eqnarray}
where $\mathcal{N}_N(U)$ is the number of microstates for $N$ indistinguishable particles with total energy $U$ fixed, and given as
\begin{eqnarray}
\label{microstates2}
%\Omega=\frac{P_N}{(2\pi)^{3N}} \frac{1}{N!}\;,
\mathcal{N}_N(U)=\frac{P_N(U)}{(2\pi)^{3N}} \frac{1}{N!}\;,
\end{eqnarray}
where $P_N(U)$ is the $N$-particle phase space, which can be expressed as
\begin{eqnarray}
\label{VPS}
P_{N}(U)=\int A^{N/2} d^{3} r_1 d^{3} r_2...d^{3} r_N \;\; d^{3} \mathrm{p}_1 d^{3} \mathrm{p}_2...d^{3} \mathrm{p}_N\; \Theta\left[U-\sum_i^N E_i (\mathbf{p}_i(\mathbf{r}_i))\sqrt{B(\mathbf{r}_i)}\right]\;,
\end{eqnarray}
for the metric given in Eq.~(\ref{metric}).  The $\sqrt{B}$ factor in the bracket accounts for the gravitational redshift and a factor of $\sqrt{A}$ is to treat spatial integral properly. The step function $\Theta$ function is included to account for the fact that the total energy of the system at the spatial infinity is fixed in the microcanonical ensemble. This function should not be confused with the one used in the one-particle phase space in the canonical ensemble, as in Eq.~(\ref{phasespace}). The latter is to find an appropriate definition for conserved energy for each particle~\cite{Padmanabhan:1989qn, Kolekar:2010py} and account for the redshift effect $E(\mathbf{r}_i)\sqrt{B(\mathbf{r}_i)}=E_\infty$. Hence it is contained in each of the $N$ integrals in Eq.~(\ref{VPS}) and not explicitly shown here.

We now proceed with an example to derive the total entropy from the fundamental relation Eq.~(\ref{Smicrostates}) and identify the corresponding thermodynamic volume. For simplicity, we focus on massless semi-classical ideal gas.
The $N$-particle phase space for this case can simply be found as
\begin{eqnarray}
\label{phasespace2}
P_N(U)&=&\frac{(8\pi)^N}{(3N)!} \;U^{3N} \prod_{i=1}^N \int \sqrt{\frac{A(r_i)}{B(r_i)^3}} d^3 r_i
= \frac{(8\pi)^N}{(3N)!} \;U^{3N} V_{th}^N\;.
\end{eqnarray}
As in the case of the canonical ensemble, we identify at this stage the thermodynamic volume as that in Eq.~(\ref{Vth0}).
%
%\begin{eqnarray}
%V_{th}&=&\int_0^R \sqrt{\frac{A(r)}{B^{3}(r)}} \;d^3r\;.
%\end{eqnarray}
% 
Then, from Eqs.~(\ref{Smicrostates}), (\ref{microstates2}), and (\ref{phasespace2}), we have
\begin{eqnarray}
\label{entropymicro}
S=\ln\mathcal{N}_N(U)= N\left( 4+\ln\left[\frac {U^3 V_{th}}{27\pi^2 N^4}\right]\right)\;.
\end{eqnarray}
where $N!\approx (N/e)^N$ is employed as usual.
From Eq.~(\ref{statefunctionsS}), the global variables are then found to satisfy
\begin{eqnarray}
\label{microUpglobal}
U=3NT_{\infty}\;,\qquad 
p_{\infty}=\frac{NT_{\infty}}{V_{th}}\;,\qquad
\mu_\infty&=&-T_{\infty} \ln\left[\frac {U^3 V_{th}}{27\pi^2 N^4}\right]\;,
\end{eqnarray}
justifying the identification of $V_{th}$. Therefore, entropy becomes 
\begin{eqnarray}
\label{microSglobal}
S&=& \ln\mathcal{N}_N(U)=N\left( 4-\frac{\mu_{\infty}}{T_{\infty}}\right)\;,\nonumber\\
&=&\int_0^R \left(4-\frac{\mu}{T}\right)n\sqrt{A}\;d^3r= \int_0^R s \sqrt{A}\;d^3r
\end{eqnarray}
in consistency with the local descriptions. Here, the local Euler equation, given in (\ref{Gibbs-Duhem}), is used for $\rho=3p$, as well as the condition that $\mu/T$ is constant. Finally, notice that thermodynamic volume $V_{th}$ found above is by default consistent with our general definition in Eq.~(\ref{dVth2}) based on entropy, given in Eq.~(\ref{entropymicro}), further confirming this identification of $V_{th}$.

%%%%%%%%%%%%%%%%%%%%%%%%%%%%%%%%%%%%%%%%%%%%%%%%%%%%%%%%%%%%%%%%%%%%
\section{Field equations for compact objects\label{sec:fieldeqs}} 
In this appendix, we provide details for field equations used in Sec.~\ref{sec:compactobj}. 
Firstly in GR, Einstein field equations for static, spherically symmetric solutions can be reduced to the momentum conservation of the stress tensor Eq.~(\ref{encons0}) and the ToV equation 
\begin{eqnarray}\label{eq:ToV}
-r^2p'(r)=G\mathcal{M}(r)\rho(r)\left[1+\frac{p(r)}{\rho(r)}\right]\left[1+\frac{4\pi r^3p(r)}{\mathcal{M}(r)}\right]\left[1-\frac{2G\mathcal{M}(r)}{r}\right]^{-1}\,,
\end{eqnarray}
where $G=\lp^2$ is the Newtonian constant and $\mathcal{M}(r)\equiv \int_0^r 4\pi r'^2\rho(r')dr'$ is the mass profile. 
For a given EoS, the profiles $p(r)$ and $\mathcal{M}(r)$ are solved simultaneously for a given value of central pressure $p_c$ at the origin. 
It turns out that these solutions have a simple scaling behavior. Defining the following dimensionless quantities
\begin{eqnarray}\label{eq:scalingToV}
\tilde{p}=p\,\lambda^4,\,
\tilde{\rho}=\rho\,\lambda^4,\,
%\tilde{k}_F=\frac{k_F}{\lambda^4},\,
\tilde{r}=r\, \lambda^{-2} \lp,\,
\tilde{\mathcal{M}}(\tilde{r})=\mathcal{M}(r) \,\lambda^{-2} \lp^3\,.
\end{eqnarray}
where $\lambda$ is some length scale. Eq.~(\ref{eq:ToV}) can be used to solve $\tilde{p}(\tilde{r}), \tilde{\mathcal{M}}(\tilde{r})$, and then solutions for $p(r), \mathcal{M}(r)$ for an arbitrary $\lambda$ can be obtained by the scaling.   

Inside the objects, $r<R$, the metric functions are given by 
\begin{eqnarray}
A(r)&=&\left[1-\frac{2G\mathcal{M}(r)}{r}\right]^{-1}\,,\nonumber\\
\frac{B(r)}{B(R)}&=&\exp\left(-\int_r^R \frac{2G}{r'^2}\left[\mathcal{M}(r')+4\pi r'^3 p(r')\right]\left[1-\frac{2G\mathcal{M}(r')}{r'}\right]^{-1} dr'\right)\,.
\end{eqnarray}
Outside the objects, $r\geq R$, it is simply the Schwarzschild solution with $B(r)=A(r)^{-1}=1-2M/r$ and the physical mass
\begin{eqnarray}
\label{GRmass}
M=\mathcal{M}(R)\equiv \int_0^R 4\pi r^2\rho(r)dr\,.
\end{eqnarray} 
It is worth mentioning that this simple expression for the physical mass is a consequence of the Einstein field equations, and shall not be expected to be valid in a general theory of gravity.

Classical quadratic gravity is used as an example of modified gravity in the paper, with the classical action below
\begin{eqnarray}\label{eq:CQG}
S_{\mathrm{CQG}}=\frac{1}{16\pi}\int d^4x\,\sqrt{-g}\left(\Mp^2R-\alpha C_{\mu\nu\alpha\beta}C^{\mu\nu\alpha\beta}+\beta R^2\right),
\end{eqnarray}
where  $\alpha, \beta$ are dimensionless couplings associated with the quadratic curvature terms.  
This is treated as the classical approximation of the renormalizable and asymptotically free quantum quadratic gravity~\cite{Stelle:1976gc}, rather than a truncation of the effective field theory. 
Since the Weyl term $C_{\mu\nu\alpha\beta}C^{\mu\nu\alpha\beta}$ softens gravitational interaction with increasing energy, quantum quadratic gravity provides a weakly coupled field theory description for gravity at high energy scale~\cite{Holdom:2016nek}. However, it brings in the problematic spin-2 ghost with mass $m_2=\Mp/\sqrt{2\alpha}$. The ghost clearly causes problems in the classical theory, but its fate at the quantum level remains under debate. Putting aside the ghost problem, the Weyl term in Eq.~(\ref{eq:CQG}) gives rise to a new type of static and spherically symmetric solutions, 2-2-holes, where the volume shrinks to zero at the origin. It is more generic than black holes and hence may serve as the endpoint of gravitational collapse if quantum quadratic gravity is the fundamental theory for gravity.  

The 2-2-hole solutions are governed by two field equations. For simplicity, we turn off the $R^2$ contribution in Eq.~(\ref{eq:CQG}), and focus on the effects of the Weyl term. For relativistic thermal gas, by implementing the momentum conservation $p(r)B(r)^2=p_\infty$, the equations are 
\begin{eqnarray}\label{eq:fieldeq1}
H_1=0,\quad H_2=8\pi \frac{A}{B^2}p_\infty\,.
\end{eqnarray}
$H_1$ and $H_2$ are functions of the metric, 
\begin{eqnarray}
H_1&=&\frac{-\Mp^2}{r^2 A^2 \left(r B'-2 B\right)}\Big[r B A' \left(r B'+4 B\right)+A \left(r^2 B'^2-2 B \left(r^2 B''+2 r B'\right)-4 B^2\right)+4 A^2 B^2\Big]\nonumber\\
H_2&=&\frac{\Mp^2}{r^2B}(B+r B'-A B)
+\frac{\Mp^2 \lambda_2^2}{4 r^4 A^3 B^3}\Big[r^2 B^2 A'^2 \left(5 B-4 r B'\right)+A^2 \Big(r^3 B'^3-3 r^2 B B'^2-4 B^3 \nonumber\\
&&\left(r A'+2\right)\Big)+A B \left(r^3 A' B'^2+2 r B B' \left(r^2 A''+r A'\right)+4 B^2 \left(r A'-r^2 A''\right)\right)+8  A^3 B^3\Big],\nonumber\\
\end{eqnarray}
where $H_1$ depends only on the Einstein term and $H_2$ includes the essential contribution from the Weyl term. 
$\lambda_2=1/m_2$ is the Compton wavelength of the spin-2 mode. 

More generally, for the stress tensor with a nonzero trace, the field equations in Eq.~(\ref{eq:fieldeq1}) become
\begin{eqnarray}\label{eq:fieldeq2}
H_1=8 \pi \,T_\mu^\mu,\quad
H_2=8 \pi \,T_2\,.
\end{eqnarray}
The right-hand sides take more complicated forms with
\begin{eqnarray}\label{eq:fieldeq2p}
T_\mu^\mu=3p-\rho,\quad
T_2= A \,p-X\frac{2B^2}{r B'-2B} T_\mu^\mu-Y \left(\frac{2B^2}{r B'-2B} T_\mu^\mu\right)'\,,
\end{eqnarray}
where $T_\mu^\mu$ denotes the trace of the stress tensor and 
\begin{eqnarray}
X&=&\frac{r B'-2 B}{48 A B^4}\frac{\lambda_2^2}{r^2} \left[r B A' \left(r B'-8 B\right)+A \left(4 B^2-7 r^2 B'^2+2 B \left(r^2 B''+8 r B'\right)\right)-4 A^2 B^2\right],\nonumber\\ 
Y&=&\frac{(r B'-2B)^2}{12 B^3}\frac{\lambda_2^2}{r^2}\, .
\end{eqnarray} 

\raggedright  
\bibliography{References_Page}{}
\bibliographystyle{apsrev4-1}

\end{document}